\documentclass[11pt,a4paper]{article}
\usepackage{jheppub}
\usepackage{amssymb,amsmath,amsfonts,mathrsfs}
\usepackage{graphicx}
\usepackage[dvipsnames]{xcolor}
\usepackage{verbatim}
\usepackage{epsfig}
\usepackage{color}
\usepackage{multirow}
\usepackage{comment}
\usepackage{datetime}
\usepackage{slashed}
\usepackage{ulem}
\usepackage{framed}
\usepackage{longtable}
\usepackage[mathscr]{euscript}
\usepackage{cancel}
\usepackage{tensor}
 \usepackage{mathtools}
  \usepackage{caption}
  \usepackage{subcaption}
 \usepackage[multiple]{footmisc}
 \usepackage{pgfplots}
 \usepackage{natbib,bibentry}
\usepackage[applemac]{inputenc}
\usepackage{tikz}
\usepackage{placeins}
\usetikzlibrary{decorations.markings}

\newcommand{\nn}{\nonumber}

\def\be{\begin{equation}}
\def\ee{\end{equation}}
\def\bea{\begin{align}}
\def\eea{\end{align}}

\def\a{\alpha}		\def\b{\beta}		\def\g{\gamma}		\def\d{\delta}
	\def\z{\zeta}					\def\q{\theta}
				\def\l{\lambda}		\def\m{\mu}
\def\n{\nu}						\def\p{\pi}			\def\r{\rho}	
\def\s{\sigma}

\title{2-charge circular fuzz-balls and their perturbations}
\author{Massimo Bianchi,}
\author{Giorgio Di Russo}

\affiliation{Dipartimento di Fisica,  Universit\`a di Roma ``Tor Vergata"  \& Sezione INFN Roma2, Via della ricerca scientifica 1, 00133, Roma, Italy}

\abstract{We perform a detailed study of perturbations around 2-charge circular fuzz-balls and compare the results with the ones obtained in the case of `small' BHs.
In addition to the photon-sphere modes that govern the prompt ring-down, we find a new branch of long-lived QNMs localised inside the photon-sphere at the (meta)stable minimum of the radial effective potential. The latter are expected to dominate late time signals in the form of `echoes'. 
Moreover, contrary to `small' BHs, some `static' tidal Love numbers are non-zero and independent of the mass, charges and angular momentum of the fuzz-ball.
We rely on the recently established connection between BH or fuzz-ball perturbation theory and quantum Seiberg-Witten curves for ${\cal N}=2$ SYM theories, which in turn are related to Liouville CFT via the AGT correspondence. We test our results against numerical results obtained with Leaver's method of continuous fractions or Breit-Wigner resonance method for direct integration and with the WKB approximation based on geodesic motion. We also exclude rotational super-radiance, due to the absence of an ergo-region, and absorption, due to the absence of a horizon.}

\begin{document}
\maketitle
\flushbottom 
\section{Introduction}

Discriminating fuzz-balls from BHs is a challenging endeavour \cite{Guo:2017jmi, Bena:2020see, Bianchi:2020bxa, Bena:2020uup, Bianchi:2020miz, Mayerson:2020tpn, Ikeda:2021uvc, Bah:2021jno, Fransen:2022jtw, Cano:2022wwo}. Recently significant progress has been made in the analysis of BH and fuzz-ball perturbation theory relying on the remarkable connection with quantum Seiberg-Witten (SW) curves for ${\cal N}=2$ SYM theories \cite{Aminov:2020yma, Bianchi:2021xpr, Bianchi:2021mft, Bonelli:2021uvf, Bonelli:2022ten, Consoli:2022eey}. Unfortunately this approach requires separability and integrability of the relevant dynamics. Since most micro-state geometries only admit time translation isometry, little progress can be made in this direction unless one restricts one's attention onto very special classes of micro-states.
Circular profiles in the two-charge system stand out as a very convenient playground to test ideas and address long-standing issues \cite{Lunin:2001jy, Mathur:2009hf}. Indeed both geodesic motion and wave equation are separable in this background and the study of Quasi Normal Modes (QNMs), absorption and tidal  deformability can be tackled with renewed strength thanks to the availability of new powerful techniques borrowed from SYM.

Moreover varying the radius $a_f$ of the circular profile one can interpolate between a `small' BH\footnote{The BH is `small' in that it has zero horizon area in the supergravity approximation.} ($a_f=0$) and  flat space-time ($a_f\rightarrow \infty$). One can envisage an average of the circular fuzz-ball observables and a comparison with the small BH ones. This would be a first step towards a much more laborious analysis for generic micro-states of `large' (BPS) BHs with at least three charges in $D=5$ or four charges in $D=4$. 

Our toy model exposes new features that in principle should allow to discriminate circular fuzz-balls from (small) BHs. As we will see the absence of a horizon, replaced by a smooth cap in the geometry,  produces a new set of long-lived QNMs that correspond to meta-stable states trapped in the `potential well' inside the photon-sphere that slowly leak out to infinity by `quantum' tunnelling through the barrier\footnote{We put `quantum' in quotes since we actually deal with 'classical' waves!}. In binary mergers these long-lived QNMs should appear in the echoes \cite{Abedi:2016hgu, Cardoso:2017cqb, Oshita:2018fqu, Wang:2018gin, Barack:2018yly, Wang:2019rcf, Cardoso:2019rvt, Barausse:2020rsu} that dominate the late stage of the GW signal and should be `easily' recognisable from the photon-sphere QNMs that dominate the prompt ring-down phase. 

Morever, despite the presence of angular momentum in the two independent spatial hyperplanes, the solution has no ergo-region ($g_{tt}<0$ always), so no (near) super-radiant modes are present in the spectrum. The absorption cross section is zero, while it is non-vanishing for the small BH. We also point out differences in the tidal deformability both in the static ($\omega=0$) and non-static ($\omega\neq 0$) case. Such a comparison between `small' BHs and `stringy' micro-states is very intriguing and, combined with analysis of chaotic behaviour in scattering amplitudes involving highly excited string states 
\cite{Gross:2021gsj, Rosenhaus:2021xhm, Firrotta:2022cku, Bianchi:2022mhs}  or in BH dynamics \cite{Shenker:2013pqa, Balasubramanian:2016ids, Saraswat:2021ong} should pave the way to further our understanding of the string-BH correspondence \cite{Horowitz:1996nw, Damour:1999aw}.

The plan of the paper is as follows.

We will start by reanalysing massive probes in the 2-charge small BH background. After discussing geodesic motion we pass to consider scalar wave equations that we initially solve in the WKB approximation. We then establish the dictionary with the quantum SW curve for ${\cal N}=2$ SYM theory with gauge group $SU(2)$ and no hypers. In turn this is equivalent to a (modified) Mathieu equation (a Doubly Reduced Doubly Confluent Heun Equation or DRDCHE for short) whose solution is analogous to the one for (massive) probes in the D3-brane background. The angular part is elementary in that it is simply given in terms of spherical harmonics. Finally we compare the results with the ones obtained using numerical methods based on Leaver's continuous fractions.   

We  then address the same issues in the D1/D5 $\approx$ D3/D3' circular fuzz-ball of radius $a_f$. After recalling the form of the metric, we study both geodesic motion and scalar wave equation and show separability. The new feature is the dependence of the separation constant $K^2$ on the combination $a_f\omega$ very much as in rotating Kerr BHs. Both for the radial and the angular dynamics the relevant ODE is a Reduced Confluent Heun Equation or RCHE for short that can be identified as the quantum SW curve for ${\cal N}=2$ SYM theory with gauge group $SU(2)$ and $N_f=(2,0)$ hypers. We study perturbations and find a new class of QNMs that are localised at the (meta)stable minimum of the radial effective potential that differ from the `standard' ones, known as the light-ring or photon-sphere modes, localised at the unstable maximum of the effective potential. In order to illustrate the analysis we consider two particular cases $\ell\sim m_\psi$ (with $m_\phi =0$) and $\ell\sim m_\phi + .. $    (with $m_\psi =0$) corresponding to `motion' in two special hyperplanes, $\theta=0$ and $\theta=\pi/2$ respectively. We check the results against direct integration and Breit-Wigner resonances, especially tailored for the long-lived modes.

We also study tidal deformability and show that some static Tidal Love Numbers (TLNs) are non-vanishing and independent of the fuzzball parameters in a certain normalisation. This is a striking result that may be taken as another smoking gun for the existence of some micro-structure at the would-be horizon scale. 

Finally we argue for the absence of instabilities or super-radiance due to the absence of an ergo-region, since $g_{tt}<0$ everywhere, and the lack of  absorption due to the absence of a horizon where to dump the incoming waves. 

We conclude with a summary of our results and an outlook.                    

Two appendices contain a brief introduction to the AGT corrspondence and a collection of tables of QNMs for fuzz-balls.

\section{2-charge systems}

The 2-charge (BPS) system is a very useful play-ground to test ideas and conjectures concerning the fate of (charged) BHs in string theory and arguably in any consistent theory of quantum gravity. 

There is a variety of ways of realising this system: from momentum-winding states to bound states of D1- and D5-branes (or equivalently D3-D3'). While the string theory description is always reliable / consistent, in fact equivalent in different frames, the validity of the low-energy supergravity approximation lead one to focus on the D-brane picture.

The degeneracy of (micro)states with fixed charges $P$ and $W$ grows as
\be
d(P,W) = \exp{4\pi\sqrt{PW}} 
\ee
even though in the supergravity description the solution has no event horizon, one can argue for the presence of a `stretched horizon'  \cite{Sen:2004dp} and for this reason the solution it is often dubbed a 'small BH' \cite{Cano:2018hut}.  Adding one more charge $K$ allows to reproduce the asymptotic behaviour of `large' (BPS) BHs in $D=5$ with   
\be
d(P,W, K) = \exp{4\pi\sqrt{KPW}} = e^{S_{BH}}
\ee
and explain the microscopic origin of the BH entropy \cite{Strominger:1996sh}. Unfortunately, despite significant progress \cite{Bena:2016ypk}, only a small fraction of the 3-charge micro-states are explicitly known \cite{Cvetic:2022fnv} and even for them probing the geometry with massless/ massive particles or waves turns out to be rather laborious due to the lack of separability/integrability with flat space-time asymptotics\footnote{With AdS boundary conditions, the $(1,0,n)$ system looks separable.} \cite{Bianchi:2018kzy, Bena:2019azk}.

For this reason, we focus on 2-charge states in $D=5$ in the D-brane formulation.
We choose the `string' profile to be circular \cite{Lunin:2001jy, Bianchi:2017sds} to achieve integrability while keeping one free parameter (the radius $a_f$) that should allow to test averaging and other effects in the relevant (reduced) ensemble of micro-states. The most general profile can oscillate in 8 transverse directions. The relevant CFT is an ${\cal N}=(4,4)$ $\sigma$-model with target space $(T^4)^N/S_N$ with $N=PW$ that admits a marginal deformations interpolating from the `free' orbifold point (at weak coupling), whereby the theory is integrable, to the `near'-horizon limit (at strong coupling), whereby holography connects the CFT to Type IIB strings on $AdS_3\times S^3\times T^4$. One can replace $T^4$ with $K3$ slightly changing the description but with similar properties.   
 
One can also interpolate between the non-singular horizonless geometry and the singular small BHs varying the parameters of the profile. In particular, for the circular profile we focus on, this corresponds to taking $a_f\rightarrow 0$. This will be crucial for our investigation and for testing (our) conjectures on how the `small' BH behaviour is reproduced by the micro-states. Moreover finding  `classical' observables that discriminate (small) BHs from 2-charge (circular) fuzz-balls may suggest that a similar situation should hold true for far more interesting large (BPS) BHs in $D=5$ (3-charge) and $D=4$ (4-charge). Hopefully one may expect a similar analysis could carry on to neutral non-BPS BHs and their micro-states \cite{Bah:2022yji}. Once again one should keep in mind that  the explicitly known solutions are not particularly amenable to the study of perturbations.  Very much as for super-strata (and even more so for multi-center solutions) they only admit reduced isometry, typically time-translation and maybe one rotation. Another interesting possibility is the breaking of equatorial symmetry \cite{Fransen:2022jtw} that is not incompatible with integrability.

 \section{BH and fuzz-ball PT: a lightning  review}
 
BH perturbation theory dates back to the early '60's when Regge, Wheeler \cite{Regge:1957td} and then Zerilli \cite{Zerilli} studied linearised wave equations around Schwarzschild geometry\footnote{According to K.~S.~Thorne \cite{KSThorne}, `singularity' at that time meant the `horizon', the original emphasys  was on 'wormhole', a term coined by J.~A.~Wheeler ... as `black-hole', or `hair, or `S-matrix'.}

Later on Teukolsky \cite{Teuko}, studied the same problem in Kerr geometry. While in the spherically  symmetric case, the angular part completely decouples from the radial part, in the rotating case, the dynamics quite surprisingly separates but the separation constant, a.k.a. Carter constant \cite{ChandraBH}, depends on the frequency of the perturbation and is not exactly known {\it i. e.} $K^2= \ell(\ell+1) + {\cal O}(a_J\omega)$ where\footnote{We denote the ratio $J/M$ by $a_J$, since we will use $a$ and $a_D$ to denote the fundamental periods of the SW curve and $a_f$ to denote the radius of the circular profile later on.} $a_J=J/M$.

\subsection{Massive scalar perturbations}

For (massive) scalar perturbations of a BH or a fuzzball, one has
$$
{\Box} \Phi = \mu_0^2 \Phi
$$
where ${\Box} = \sqrt{||g||}^{-1} \partial_\mu (\sqrt{||g||} g^{\mu\nu} \partial_\mu\Phi)$ is the scalar wave operator and $\mu_0 = 2\pi/\lambda$ is its inverse Compton wave-length\footnote{One may object that the notion of Compton wave-length $\lambda = \hbar /mc$ involves Planck constant which looks odd for `classical' waves. Yet for KK excitations $\lambda \approx 2\pi R/n$ and $\mu\approx n/R$, since $\hbar$ cancels.}. Setting 
$$
\Phi = e^{-i\omega t} e^{im_\phi\phi} R(r) S(\theta)
$$
allows in all the cases of interest here to separate the dynamics.

In particular for Schwarzschild BHs one finds \cite{ChandraBH}
$$
 R''(r) + {2(r-M)\over r(r-2M)} R'(r)+ \Bigg[{r^2\omega^2\over(r-2M)^2}-{\ell(\ell+1)\over r(r-2M)}\Bigg]R(r)=0
 $$ 
 and
 $$
{1\over\sin\q}\partial_\q\Big[\sin\q\partial_\q S(\q)\Big]-{m_\phi^2\over\sin^2\q}S(\q)=-K^2S(\q)
 $$
 where $K^2 = K_0^2 = \ell(\ell+1)$ and $S_{\ell,m_\phi}(\theta)e^{im_\phi\phi} = Y_{\ell,m_\phi}(\theta, \phi)$ are the familiar spherical harmonics. 

Both radial and angular equations can be put in canonical Schr\"odinger-like form
\be 
\Phi''(y) + Q_{BH}(y) \Phi(y) = 0
\ee

Notice that for the radial equation one has   
$$
Q_{r,w}(r) = f(r) [ \omega^2 - V_{BH}(r)]
$$
where $f(r)= \left(1+{1\over{r\over2M}-1}\right)^2$ plays the role of an effective/running mass and $V_{BH}(r)$ depends on $\ell$ but neither on $\omega$ nor on $m_\phi$ due to spherical symmetry. 

For rotating BHs and fuzz-balls $V_{eff}$ may depend on $\omega$ both explicitly and implicitly through the dependence of the `conserved' quantities (like $K^2$) on  $\omega$.

In order to reduce the mass to a constant (one), a new radial variable $\tilde{r}$, known as `tortoise' coordinate, can be introduced that satisfies 
$$
d\tilde{r} = \sqrt{f(r)} dr 
$$
in this way the `horizon', if present, is mapped to $\tilde{r} = -\infty$ while at spatial infinity $\tilde{r} \approx r$. For Schwarzschild BHs, the `tortoise'  coordinate coincides with the time for radial fall, {\it viz.}
$$\tilde{r} = r + 2M\log\left({r\over2M}-1\right) = t_{rad-fall} $$ 
Bringing the resulting equation into canonical form introduces further terms in the effective potential
$$
\tilde{V}_{BH}(\tilde{r}) = V_{BH}(r(\tilde{r})) + \Delta V_{BH}(\tilde{r})
$$
For Schwarzschild BHs one has the Regge-Wheeler (Zerilli) potential \cite{ChandraBH}
$$
\tilde{V}_{RWZ}(\tilde{r}) = {(r-2M)\Big[\ell(\ell+1)r+2M\Big]\over r^4}
$$
For rotating (Kerr-Newman) BHs  the separation constant  $K^2= \ell(\ell+1) + {\cal O}(a_J^2\omega^2$) depends on $\omega$ and $a_J=J/M$ and the definition of the tortoise coordinate is ambiguous.

For non-extremal geometries both angular and radial equations boil down to Confluent Heun Equations (CHE) with two regular singular points (at the horizons) and an irregular singular point (at infinity). 

Until recently, neither solutions nor connection formulae for Heun Equation and its cousins were known in closed form. As a consequence the most efficient approach to the solution with prescribed boundary conditions (needed {\it e.g.} for QNMs) was based on numerical methods developed by Leaver \cite{Leaver1}, relying on continuous fractions that allow to solve (numerically) 3-term recursion relations. When these involve more than 3 terms, Gaussian elimination may be employed to subsequently reduce the number of terms \cite{Leaver2}. A relatively good estimate  is provided by the WKB approximation that connects with geodesic motion in the eikonal limit (for large $\ell$) that can be used as a `seed' for numerics. 

In particular, QNMs \cite{Iyer:1986vv} can be related to a special class of geodesics representing closed (un)stable (`circular') orbits around the compact object (BH / fuzz-ball). They are referred to as `critical' in that they are  determined by the simultaneous vanishing of the radial momentum $P_r^2 = Q(r) $ and its derivative (`zero force'):
\be
Q(r)=Q'(r)=0
\ee

Note that the zero-force condition is compatible with both a maximum (classically unstable) and a minimum (classically stable). When replacing `particles' with waves, the minimum may become meta-stable in that the wave may tunnel the potential barrier represented by the photon-sphere and leak out to infinity. 

This means that one should expect (at least) two sets of QNMs for Exotic compact objects (ECOs) or fuzz-balls. For massive probes there might be (innermost) stable circular orbits (ISCO) where light particles can be trapped and emit electro-magnetic radiation that 'pollutes' (in that it represent an environmental effect) the photon-sphere and other features of the BH {\it in vacuo} \cite{EventHorizonTelescope:2019dse,Johnson:2019ljv,Kapec:2022dvc,Bianchi:2020yzr}. 

The photon-sphere modes, that dominate the prompt ring-down of the remnant of a binary merger, correspond to unstable modes localised around the tip of the potential. These are the `standard' ones that are present also in the small BH limit at $a_f=0$ as we will see. The two conditions allow to determine the critical radius $r_c$ and the critical impact parameter $b_c= (K/E)_c$ in terms of the other parameters. 

In the `classically-allowed' regions where $Q_{r,w}$ is positive and large\footnote{We denote by $Q_{r,w}$ the square radial momentum in the wave equation to distinguish it from $Q_{r,geo}$.}, the wave equation can be solved in a semiclassical approximation:
\be
\psi(r)={1\over\sqrt[4]{Q_{r,w}(r)}} e^{\pm i \int^r\sqrt{Q_{r,w}(r')}dr'} 
\ee
This approximation fails near the zeros of $Q_{r,w}$ called $r_\pm$ where the (radial) momentum vanishes and should be amended in the `classically-forbidden' regions where $Q_{r,w}$ is negative. The matching of the solution at two consecutive inversion points implies the Bohr-Sommerfeld (BS) quantization condition:
\be\label{wkb1}
\int_{r_-}^{r_+}\sqrt{Q_{r,w}(r)}dr=\pi \left(n+{1\over2}\right)
\ee
with $n$ a non negative integer a.k.a. overtone number. This leads to standard energy quantisation or, in our case where everything is classical, to the complex frequency of the (meta)stable modes localised at a local minimum of the potential, if present in the form of an outer or internal stable light-ring \cite{Bianchi:2020yzr}.

When the two inversion points are the ends of a classically-forbidden region and (almost) coincide (for critical geodesics), one should `analytically' continue BS formula and the integral can be approximated as follows:
\be
\label{wkb2}
\int_{r_-}^{r_+}\sqrt{Q_{r,w}(r)}dr\sim\int_{r_-}^{r_+}\sqrt{Q_{r,w}(r_c)+{Q''_{r,w}(r_c)\over2}(r-r_c)^2}dr\sim{i\pi Q_{r,w}(r_c)\over \sqrt{2 Q''_{r,w}(r_c)}}
\ee
since $Q'_{r,w}(r_c)=0$. The resulting frequencies acquire an imaginary part, i.e. $\omega=\omega_R+i \omega_I$  and characterise what are called quasi-normal modes (QNMs). 

In the WKB approximation one finds 
\be
\omega_{QNM}^{WKB} = \omega_c(\ell, ...) - i (2n+1)\lambda
\ee
where $\omega_c$ is the angular velocity of the un-stable circular orbits and $\lambda$ is the Lyapunov exponent governing the chaotic behaviour of near-critical geodesics around the photon-sphere. It can be determined in a variety of ways, the simplest being probably using `radial' geodesic deviation around $r=r_c$ 
\be
{dr\over dt} \approx \lambda (r-r_c) 
\ee
Notice that the bound on chaos $\lambda \le 2\pi K_B T_{BH}/\hbar$ set by \cite{Maldacena:2015waa} can be violated near extremality  and must be replaced with a more general bound \cite{Bianchi:2020des, Bianchi:2020yzr}.

In addition to the unstable orbits, some BHs admit stable circular orbits for massive probes outside the photon-sphere that are known as ISCOs (Innermost Stable Circular Orbits) \cite{ChandraBH}. The image produced by the ETH represents a combination of photon-sphere or photon-halo, in the case of rotating objects since $r_c$ depends on the value the angular momentum $J$ and $J_z$ of the probe, and emission from plasma populating the ISCOs, that depend on the angular momentum $J$ and $J_z$ of the probe, too \cite{EventHorizonTelescope:2019dse,Johnson:2019ljv,Kapec:2022dvc,Bianchi:2020yzr}.

Moreover, non-singular ECOs such as fuzz-balls or generic micro-states may also admit (meta)stable orbits inside the photon-sphere. 
A set of long-lived metastable modes localised around the relative minimum may arise. 

Assuming for simplicity that the radial equation can be put in the form 
\be
\psi'' + f( \omega^2 - V_{BH}) \psi =0
\ee
with $V_{BH}$ the radial effective potential. Neglecting the (mild) dependence of $V_{BH}$ on $\omega$\footnote{Otherwise one has to define $\Omega_{\pm}$:  $\omega^2 - V_{BH}(..., \omega) = (\omega - \Omega_{+})(\omega - \Omega_{-})$.} and treating $f$ as an `effective' mass, rather than introducing a tortoise radial coordinate $d\xi = \sqrt{f(r)} dr$, in the WKB approximation one finds
\be
\omega_{QNM}^{meta} = \omega^{meta}_R - i \omega^{meta}_I
\ee
with 
\be
(\omega^{meta}_R)^2 = V_{BH}(r_{min}) + \left(n+{1\over 2}\right) \sqrt{2 V_{BH}''(r_{min})\over f(r_{min})} 
\ee
and 
\be
\omega^{meta}_I = {\pi\over {\cal T}} |T|^2
\ee
where 
$$
|T|^2 = \exp\Bigg[ -2 \int_{r_-}^{r_+} dr \sqrt{f(r) (V_{BH}(r) -\omega^2)}\Bigg]
$$ 
is the tiny probability of tunnelling `under' the photon-sphere and 
$$
{\cal T}= {2\pi\over\omega_R} = {d\over v}
$$
gives an estimate of the time in between two bounces on the `wall' of width $d\approx {r_+}-{r_-}$. We will show that this indeed happens for circular fuzz-balls when $a_f\underset \sim < L$ is large enough. For small $a_f<<L$, the effective potential is too steep to accommodate any level at all.

\subsection{Other observables}

Other observables, such as the absorption cross-section, super-radiance amplification factor, echoes and tidal deformability require detailed knowledge of the connection formulae that can be derived using the remarkable correspondence between BH/fuzz-ball PT and quantum SW curves for ${\cal N}=2$ theories with $SU(2)$ gauge group and $N_f\le 4$ fundamental flavours \cite{Aminov:2020yma, Bianchi:2021xpr, Bianchi:2021mft, Bonelli:2021uvf, Bonelli:2022ten, Consoli:2022eey}. In fact the AGT correspondence (after Alday, Gaiotto, Tachikawa \cite{Alday:2009aq}) allows to gain further insights \cite{Bianchi:2021xpr, Bianchi:2021mft, Bonelli:2021uvf, Bonelli:2022ten, Consoli:2022eey} as we will see later on and in appendix \ref{AGT}.

In canonical form and for real $\omega$ one has $Q=Q^*$ so that the system admits a conserved `current' \footnote{{\it Caveat} in Liouville CFT ${k}$'s, ${p}$'s are usually purely imaginary! \cite{Bonelli:2021uvf, Bonelli:2022ten, Consoli:2022eey}.} 
$${\cal J}= {\rm Im} [\Psi^* \partial_z \Psi] = k_H (|C^{(H)}_+|^2-|C^{(H)}_-|^2) =k_\infty (|C^{(\infty)}_+|^2-|C^{(2)}_-|^2)$$   
$$
{\cal J}_{abs} = - k_H |C^{(H)}_-|^2 \quad , \quad  {\cal J}_{ref} = + k_H |C^{(H)}_+|^2
$$
$$
{\cal J}_{in} = - k_\infty |C^{(\infty)}_-|^2 \quad , \quad {\cal J}_{out} = + k_\infty |C^{(\infty)}_+|^2
$$

The absorption probability is defined by
$$
{\cal P}_{abs}(\omega) = {|{\cal J}_{abs}(\omega)| \over  |{\cal J}_{in}(\omega)|} = { k_H |C^{(H)}_-|^2 \over k_\infty |C^{(\infty)}_-|^2}  = {|C^{(\infty)}_-|^2 - |C^{(\infty)}_+|^2 \over |C^{(\infty)}_-|^2} = 1 - {|C^{(\infty)}_+|^2 \over |C^{(\infty)}_-|^2}
$$
where the latter form relies on current conservation.

Echoes are present when the reflectivity ${\cal R}$ at the would-be horizon (`cap') is nonzero ${\cal R}\neq 0$. 

 
On the other hand, in the presence of an ergo-region, one can compute the super-radiance amplification factor 
$$
Z(\omega) = {|{\cal J}_{out}(\omega)| \over  |{\cal J}_{in}(\omega)|} - 1 = { k_H |C^{(H)}_-|^2 \over k_\infty |C^{(\infty)}_-|^2}
$$
that is positive below the super-radiant threshold 
$$
\omega_{SR} \le m_\phi \Omega_H 
$$
where $\Omega_H$ is the angular velocity at the horizon and $m_\phi$ is the azimuthal number. Super-radiance amplifies the intensity of a wave  and allows to extract energy and angular momentum from a rotating or charged BH \cite{Brito:2015oca} very much as Penrose process \cite{Penrose:1971uk} uses particles to perform the same task, even in smooth horizonless geometries with an ergo-region \cite{Jejjala:2005yu, Cardoso:2005gj, Bianchi:2019lmi, Chakrabarty:2019ujg}. The end-point of this ergo-region instability is expected to be a smooth horizonless BPS geometries \cite{Giusto:2004ip}.

The small imaginary parts of the super-radiant modes of near extremal rotating BHs can be computed by means of the QNM-qSW correspondence \cite{Bianchi:2022wku}. Quite remarkably at extremality, one can also take advantage of generalised symmetry under Couch-Torrence conformal inversions \cite{CouchTorr, Cvetic:2020kwf, Cvetic:2021lss, Bianchi:2021yqs, Bianchi:2022wku} 
that exchanges horizon with infinity and  keeps the photon-sphere/halo fixed! \cite{Bianchi:2021yqs, Bianchi:2022wku}

\subsection{BH/fuzz-ball perturbations and quantum Seiberg-Witten curves}

We will now briefly describe how BH/fuzz-ball perturbations can be studied by means  of quantum SW curves of ${\cal N}=2$ SYM \cite{ SW9407087, SW9408099}. We will rely on the Hanany-Witten \cite{Hanany:1996ie} setup in which (linear quiver) theories are realised on D4-branes suspended in between NS5-branes. For our purposes it is sufficient to consider $N_c=2$ {\it i.e.} $SU(2)$ gauge group, after $U(1)$ decoupling, and $N_f\le 4$, in fact $N_L, N_R\le 2$. The latter correspond to `flavour' D4-brane infinitely extended to the Left or to the Right of the two NS5-branes \cite{Hanany:1996ie}.

In particular the CHE that encompasses all BH perturbations in $D=4$ can be shown to correspond to $N_f=3=(2,1)$ or $N_f=2=(1,1)$ for the radial equation in the extremal case (a.k.a. DCHE i.e. Doubly CHE) \cite{Aminov:2020yma, Bianchi:2021xpr, Bianchi:2021mft, Bonelli:2021uvf, Bonelli:2022ten, Consoli:2022eey, BACCP1702.01016, BACCP1812.08921}. For circular fuzz-balls both radial and angular equations, coupled to one another by Carter constant, boil down to RCHE that can be shown to correspond to $N_f=(2,0)$, reducing to  $N_f=(0,0)$ DRDCHE for the radial equation in the small BH limit  whereby $a_f\rightarrow 0$.  

The `classical' SW curve \cite{SW9407087, SW9408099}, whose (fundamental) periods $a$ and $a_D$ encode the gauge-invariant Coulomb branch parameter $$u =\langle {\rm Tr}\phi^2\rangle = a^2 + ... $$ can be written as  
\be
q y^2 \prod_{f=1}^{N_L} (x-m_f) + y \prod_{u=1}^{2} (x-e_r) + \prod_{f=1}^{N_R} (x-m_f) = 0
\ee
where $N_c=2$ is the number of colour branes, $N_L+N_R=N_f$ is the number of fundamental hypers (`flavour' branes), $q=\Lambda^\beta$ with $\beta=4-N_f$ the one-loop $\b$-function coefficient and $m_f$ are the masses of the hypers.

The quantum version obtains after turning on an $\Omega$ background \`a la Nekrasov-Shatashvili (NS) $(\varepsilon_1=\hbar, \varepsilon_2=0)$ \cite{
Marshakov:2009gn,Nekrasov:2009rc,Mironov:2009uv,Z1103.4843,BF1711.07935,GM1806.01407,GGM1908.07065}. As a result
\be 
y\rightarrow \hat{y} = y \quad , \quad x\rightarrow \hat{x} = \hbar y{d\over dy}
\ee
In this way one obtains a second order differential equation 
$$
\left[A(y) \hat{x}^2 + B(y) \hat{x} + C(y)\right] U(y) = 0
$$
with (ir)regular singularities determined by the choice of $q$, $u$ and $m_f$ that can be put in canonical form 
\be
\Psi''(y) + Q_{SW}(y) \Psi(y) = 0
\ee
or in other standard forms, depending on the number and nature of the singularities.

After `quantum' deformation, the SW differential reads
\be
\lambda_+(x)=-{x\over 2\pi i}{1\over W(x)}{dW(x)\over dx}
\ee
where $W(x)$ can obtained recursively, using the difference equation thet follows from $\hat{y}=e^{-\hbar\partial_x}$
\be
q P_L(x-{\hbar\over 2}) P_R(x+{\hbar\over 2}) W(x)W(x-\hbar) + P_0(x) W(x) + 1 =0
\ee
with 
\be
P_L(x) = \prod_{f=1}^{N_L} (x-m_f) \quad, \quad P_R(x)= \prod_{f=1}^{N_R} (x-m_f)
\quad , \quad  P_0(x) = \prod_{u=1}^{2} (x-e_r) 
\ee
For $SU(2)$ (rather than $U(2)$) one sets $e_1+e_2=0$ and $e_1e_2=-u$ (the gauge invariant Coulomb branch parameter). The quantum deformed basic period is given by
\be
a(u,q,m_f; \hbar) = \oint \lambda_+(x) dx 
\ee

Quite remarkably for $N_f=4$ ({\it i.e.} $N_L=N_R=2$) one gets an equation with 4 regular singular points that can be written in the form of Heun Equation (HE) and governs scalar perturbations\footnote{Dual to operators of dimension $\Delta=1,2$ on the boundary.} with `mass' ${\mu}^2L^2=-2$ around Kerr-Newman BHs in AdS$_4$. In order to study BH/fuzz-ball PT with flat asymptotics, one can decouple flavours one after the other by performing the double scaling limit 
\be
q_{N_f}\rightarrow 0 \quad, \quad  m_{M_f}\rightarrow \infty \quad {\rm with} \quad  q_{N_f} m_{N_f}  = q_{N_f{-}1}  \quad {\rm fixed}
\ee

In the spherically symmetric case the angular equation decouples completely from the radial equations and the `separation' constant is simply $K^2 =K_0^2=\ell(\ell+{D-3})$. 

When the BH/ECO/fuzz-ball  carries spin (e.g. $J=Ma_J$ for Kerr) the separation constant depends on the combination $a_J\omega$ that intertwines angular and radial equations. For the angular dynamics, in the qSW approach, the relevant quantisation condition involves the $a$-cycle {\it viz.}
\be
a^\theta = {\ell \over {D-3}} + {1\over 2} 
\ee
in units of $\hbar =\varepsilon_1 (=1)$, while $\varepsilon_2=0$. Since the dictionary yields
\be 
u^\theta = {1\over 4} (1+K^2) 
\ee
knowledge of $u(a)= a^2 + ...$ from the quantum Matone relation \cite{Matone:1995rx, Flume:2004rp}
\be\label{quantMat}
 u = -q  {\partial{\cal F}\over \partial q} ,
 \ee
where ${\cal F}(a;m_f;q; \hbar)$ is the quantum-deformed analytic pre-potential \cite{Nekrasov:2009rc, MM0910.5670, Z1103.4843},  is tantamount to knowledge of $K^2$ as a function of the principal quantum number $\ell$, of the `azimuthal' quantum number(s), related to the masses  $m^\theta_f$,  as well as  of $q_\theta\approx a_J\omega$. In this sense, using qSW curves \`a la NS solves completely the problem of finding the so-called spheroidal harmonics and their eigen-values  $K^2$\cite{BCC0511111}!

The analysis of the radial equation is significantly more laborious since the relevant quantisation condition involves the $a_D$-cycle {\it viz.} 
\be
2\pi i a_D = - {\partial{\cal F}\over \partial a} = - 2 a \log q + ...  = 2\pi i (n + \delta)
\ee 
where ${\cal F}(a, m_f, q, \hbar)$ is the NS prepotential, while $n$ is known as the overtone number and $\delta$ is a (mass-dependent) shift. This follows from the identification of the `vanishing' cycle in the classical limit. For the prompt ring-down QNMs the `vanishing' cycle corresponds to the photon-sphere. For the long-lived QNMs the `vanishing' cycle corresponds to the `internal' classically-stable circular orbit.

In practice one has to find the dictionary 
\be 
u= U(\ell,m_f, ... Q, M, a_J, \omega )
\ee
and invert the relevant quantum Matone relation (\ref{quantMat}) in order to express $a$ as a function of $u$ and find ${\cal F}(a;m_f;q; \hbar)$ (up to the one-loop term, that can be computed by other means \cite{Nekrasov:2009rc, MM0910.5670, Z1103.4843}). Finally one has to impose the quantisation on $a_D$ as an `eigenvalue' equation for $\omega$, whose solutions are functions of $n$, $\ell$, $m$'s, $M$, $Q$, $J$'s.  In \cite{Bianchi:2021xpr, Bianchi:2021mft, Bianchi:2022wku} we checked agreement among qSW, Leaver and WKB for a wide variety of BHs and (bound-states of) D-branes. 
 
Momentarily we will extend the analysis to the simplest (BPS) fuzz-ball, the one with circular profile in $D=5$. Although the BH limit has vanishing horizon area the degeneracy of the micro-states is a signal of a non-trivial structure at the would-be horizon that we would like to study in detail looking for new signatures and for guidance on how to perform ensemble averages.
  
As in \cite{Bianchi:2021xpr, Bianchi:2021mft, Bianchi:2021yqs, Bianchi:2022wku}, we focus on scalar perturbations, extension to spin $s=1,2$ is straightforward but more technical and tedious\footnote{Since the backgrounds we consider are BPS, in principle one can relate linearised scalar fluctuations to vector and tensor ones by means of the unbroken supersymmetry.}.

\section{Probing and re-probing small BHs with massive particles and waves }

In order to set the stage for our investigation and for later comparison, we will start by discussing (massive) neutral scalar perturbations of the `small' BH solution of the D1/D5 branes system. Sometimes we switch to the T-dual D3/D3' description that makes more transparent the decoupling of the $T^4$ part of the geometry for neutral probes ({\it i.e.} those carrying only momentum along $S^1$ and no KK momentum on $T^4$).

The metric (in the `democratic' D3/D3' frame) is \cite{Lunin:2001jy, Mathur:2009hf}:
\be\label{d1d5massmetric}
ds^2{=}{-}H^{-1}(dt^2{-}dz^2){+}H\Big[dr^2{+}r^2 (d\q^2{+}\cos^2\q d\psi^2{+}\sin^2\q d\phi^2)\Big]{+}\sqrt{H_1\over H_5}dz^2_{1,2}{+}\sqrt{H_5\over H_1}dz^2_{3,4}
\ee
where 
\be
H(r)=\sqrt{H_1(r)H_5(r)}\quad,\quad H_i(r)=1+{L_i^2\over r^2}
\ee
where $L_1^2 =32\pi^2 g_s N_1(\a')^3 $ and $L_5^2=g_s N \a'$. Other Type IIB supergravity fields acquire non-trivial profiles, but they will play no role in our present analysis. 

\subsection{ Massive Geodesics}

The mass shell condition for massive neutral particles in Hamiltonian formalism reads \cite{Chervonyi:2013eja, Bianchi:2017sds}:
\be \mathcal{H}= {1\over 2} g^{\mu\nu} P_\mu P_\nu = - {1\over 2}\m_0^2
\ee
with 
\be
2\mathcal{H}= -H\mathcal{E}^2+H^{-1}\Big[ P_r^2+{P_\q^2\over r^2}+{J_\psi^2\over r^2\cos^2\q}+{J_\phi^2\over r^2\sin^2\q}\Big]+\sqrt{H_5\over H_1}\left(P_1^2{+}P_2^2\right)+\sqrt{H_1\over H_5}\left(P_3^2{+}P_4^2\right)
\ee
where $\mathcal{E}^2=E^2-P_z^2$ and the conserved momenta are:
\be
P_t={\partial \mathcal{L}\over \partial \dot{t}}=-E\quad,\quad P_z={\partial \mathcal{L}\over \partial \dot{z}}\quad,\quad P_\psi={\partial \mathcal{L}\over \partial \dot{\psi}}=J_\psi\quad,\quad P_\phi={\partial \mathcal{L}\over \partial \dot{\phi}}=J_\psi\quad,\quad P_I={\partial \mathcal{L}\over \partial \dot{Z^I}}
\ee
The problem is separable and the remaining dynamical momenta are
\begin{align}
&P_r^2=Q_{r,geo}(r)=H_1H_5\mathcal{E}^2-{K^2\over r^2}-H_5\left(P_1^2+P_2^2\right)-H_1\left(P_3^2+P_4^2\right)-H\m_0^2\\\nn
&P_\q^2=Q_{\q,geo}(\q)=K^2-{J_\psi^2\over \cos\q^2}-{J_\phi^2\over \sin^2\q}
\end{align}
where $K^2$ is the separation constant, associated to the total angular momentum. For later use one can define an impact parameter $b=K/\mathcal{E}$. 

Due to spherical symmetry, geodesics are planar. 
For large $b$ the geodesics are open and correspond to small deflections angles. For small $b$ the geodesics end on the horizon $r=0$. For a critical value $b=b_c$ the geodesics wind around the small BH and end on a limiting cycle $r=r_c$, that represents a critical unstable orbit. 
In order to determine  $(r_c,b_c)$ or equivalently $(r_c,K_c)$ one has to solve the critical conditions $Q_{r,geo}(r_c,K_c)=Q'_{r,geo}(r_c,K_c)=0$.

For simplicity, we will take $L_1=L_5=L$, that allow to separate the dynamics even for massive probes. Solving the criticality conditions, one finds
\begin{align}\label{massived1d5crit}
&r^2_c={L^2 {\mu}^2+K_c^2-2L^2 \mathcal{E}^2\pm\sqrt{(2L^2 \mathcal{E}^2-L^2{\mu}^2-K_c^2)^2-4L^4\mathcal{E}^2(\mathcal{E}^2-{\mu}^2)}\over 2(\mathcal{E}^2-{\mu}^2)}\\\nn
&(2L^2 \mathcal{E}^2-L^2{\mu}^2-K_c^2)^2=4L^4\mathcal{E}^2(\mathcal{E}^2-{\mu}^2)
\end{align}
where $\mu^2=\mu_0^2+\sum_i P_i^2$. The only acceptable solution of the previous system is:
\be
K_c=L\left(\mathcal{E}+\sqrt{\mathcal{E}^2-{\mu}^2}\right)\quad,\quad r_c=L\sqrt[4]{{\mathcal{E}^2\over\mathcal{E}^2-{\mu}^2}}
\ee
that corresponds to the un-stable orbits forming the photon-sphere (or light-ring) when $\mu=0$. Notice that there are neither stable circular orbits (ISCOs) outside nor inside the photon-sphere, contrary to the 4-d case. While no large radius ISCOs are expected since for large $r$ the potential behaves as $1/r^2$, that does not admit such stable orbits, there are none even at `small' $r$ where relativistic corrections become important. As a result we do not expect any additional (long-lived / meta-stable) QMNs beyond the `standard' prompt ring-down modes.

In order to address this and related issues we will now consider wave propagation and scattering in the background of a small BH.

\subsection{Wave equation}
The wave equation for massive scalar probe in the D1-D5 background \eqref{d1d5massmetric} is of the form:
\be
(\Box-\mu_0^2)\Phi=0
\ee
The ansatz
\be
\Phi=\exp\left(-i \omega t+i P_z z+i m_\phi \phi+i m_\psi \psi+i\sum_{I=1}^4P_I Z^I\right)R(r) S(\q).
\ee
allows to separate the dynamics. The (polar) angular equation 
\be
\Bigg[{1\over \sin2\q}\partial_\q (\sin2\q \partial_\q)-{m_\psi^2\over \cos^2\q}-{m_\phi^2\over \sin^2\q}\Bigg]S(\q)=-\ell(\ell+2) S(\q)
\ee
is solved by the spherical harmonics on $S^3$ which are related to the characters of $SU(2)$ \cite{Bianchi:2017sds}. 
The radial equation becomes:
\be
R''(r){+}{3\over r} R'(r){+}\Bigg[H^2 \tilde{\omega}^2{-}{\ell(\ell+2)\over r^2}{-}H_5(P_1^2{+}P_2^2){-}H_1(P_3^2{+}P_4^2){-}\mu_0^2H \Bigg]R(r){=}0
\ee
where $\tilde{\omega}^2=\omega^2-P_z^2$. As for the geodesics, we will set $L_1=L_5=L$ from now on, without much loss of generality. The resulting radial equation can be put in canonical form setting $R(r)=r^{-3/2}\Psi(r)$:
\begin{align}\label{seq}
&\Psi''(r)+Q_{r,w}(r)\Psi(r)=0\\\nn
&Q_{r,w}(r)={(\tilde{\omega}^2-{\mu}^2)r^4+\Big[2L^2\tilde{\omega}^2-L^2 {\mu}^2+{1\over 4}-(\ell+1)^2\Big]r^2+L^4\tilde{\omega}^2\over r^4}
\end{align}
\subsection{WKB approximation}
 
Eq \eqref{seq} can be viewed as the Schr\"odinger equation for a non-relativistic particle with an `effective mass' $m_{eff} = f/2$ and  `energy' ${\cal E}= \omega^2$ in a potential {\it i.e.} 
\be
Q_{r,w} = f (\omega^2 - U) = 2m_{eff} ({\cal E} - V)
\ee 

The `effective mass' can be set to unity by the introduction of a `tortoise' coordinate $\xi$ such that $f= 2m_{eff} = (d\xi/dr)^2$ and a further redefinition of $\Psi$ necessary to bring the equation back to canonical form.

As already mentioned, there are several observables that can be extracted from  solving the wave equation. In particular the QNMs correspond to outgoing waves at infinity that satisfying ingoing boundary conditions at the (zero-area) horizon $r=0$.

In the WKB approximation, it is reasonable to assume that QNMs satisfy $|\omega_I|\ll |\omega_R|$ so that  \eqref{wkb1} can be solved perturbatively in $\omega_I$. At zeroth order, we compute $\omega_R\approx \omega_c(\ell, ...) $ and plugging the expression of $\omega_R$ in (\ref{wkb2}) we find:
\be
\omega^{QNM}_{WKB}={(\ell+1)^2-{1\over4}+L^2{\mu}^2\over2L\sqrt{(\ell+1)^2-{1\over4}}}-i (2n+1){\Big[(\ell+1)^2-{1\over4}-L^2{\mu}^2\Big]^{3\over2}\Big[(\ell+1)^2-{1\over4}+L^2{\mu}^2\Big]^{1\over2}\over 4L\Big[(\ell+1)^2-{1\over4}\Big]^2}
\ee

As already observed, the imaginary part is related to the Lyapunov exponent governing the chaotic behaviour of nearly critical geodesics around the light-ring (photon-sphere). The numerical values of QNM in WKB approximation for the lowest overtone ($n=0$) and for various $\ell$ are displayed in the first column of Table \ref{tabBHQNM}.

\subsection{Massive D1/D5-SW-Mathieu dictionary}
The radial wave equation \eqref{seq} for the massive scalar field on a D1/D5 background is {\it mutatis mutandis} the same as on a D3 background \cite{Gubser:1998iu}, that can be mapped to the quantum SW curve of pure SU(2) gauge theory, or equivalently to the (modified) Mathieu equation, as shown in \cite{Bianchi:2021xpr, Bianchi:2021mft}. 
Setting $$r={\hbar \tilde{\omega}L^2\over2}\sqrt{y}=L\sqrt[4]{{\tilde{\omega}^2\over\tilde{\omega}^2-{\mu}^2}}e^{iz}$$
the relevant potentials are:
\begin{align}
&Q_{SW}(y)={4qy^2+(\hbar^2-4u)y+4\over 4\hbar^2 y^3}\\\nn
&Q_M(z)=\a-2\b \cos(2z)
\end{align}
The gauge/gravity/Mathieu dictionary reads:
\be\label{mathieudict}
\a={4u\over\hbar^2}=(\ell+1)^2+L^2 {\mu}^2-2L^2\tilde{\omega}^2\quad,\quad\b={4\sqrt{q}\over\hbar^2}=\tilde{\omega} L^2\sqrt{\tilde{\omega}^2-{\mu}^2}
\ee

In the Mathieu `version' $\alpha$ plays the role of energy eigen-value and the symmetry $z\leftrightarrow -z$ is nothing but the generalised Couch-Torrence symmetry $r\leftrightarrow  r^2_c/r$ \cite{CouchTorr} that exchanges infinity and horizon keeping the photon sphere fixed. As a result Mathieu equation admits fundamental even and odd solutions. 

Moreover the potential is periodic under $z\rightarrow z+\pi$, so that a non-trivial solution of Mathieu equation $w(z)$ is quasi-periodic:
\be
w(z+\p)=e^{\p i \n}w(z)
\ee
with $\n(\a,\b)$ known as Floquet exponent, that in turn can be related to the quantum $a$-period at weak coupling and to the quantum $a_D$-period at strong coupling. At weak coupling, one finds:
\be
a(u,q)={\hbar\over 2}\nu(\a,\b)
\ee
In the weak coupling limit ($\b\ll\a$) the energy eigenvalue is:
\begin{align}
\label{ufroma}
&\a=\n^2+{\b^2\over2(\n^2-1)}+{(5\n^2+7)\b^4\over 32(\n^2-1)^3(\n^2-4)}+...\\\nn
& u = a^2 + {2 \over(4a^2-1)}q + {7+20a^2\over2(a^2-1)(4a^2-1)^3}q^2+...
\end{align}
 Inverting this relation {\it i.e.} finding $\nu = 2 a$ in terms of $u $ is subtle. One has to pay particular attention when -- as in the case at hand -- $\alpha=4u\approx n^2 = \nu^2 + ...$ is close to a square integer, so that $\nu = n + \delta n(q)$. The standard formula does not work since it is derived with the tacit assumption that the poles of the quantum SW differential $\lambda_+(x)$ be different from one another. When two poles coalesce, the residues must be computed with extra care and the expression of $a(u)$ acquires a different form.  
     
The photon-sphere can be more conveniently described  in the opposite limit $\a\sim\pm2\b$ ($u\sim \pm2\sqrt{q}$) that corresponds to the strong coupling regime of the gauge theory where the $a_D$-cycle degenerates. In this limit the energy eigenvalue appears:
\begin{align}
&\a=2 \beta+2s\sqrt{-\beta } -\frac{s^2+1}{8} -\frac{s^3+3 s}{128 \sqrt{-\beta }}+\frac{5 s^4+34 s^2+9}{4096 \beta }-\frac{33
   s^5+410 s^3+405 s}{131072 (-\beta )^{3/2}}+\\\nn
&-\frac{63 s^6+1260 s^4+2943 s^2+486}{1048576 \beta ^2}-\frac{527 s^7+15617 s^5+69001 s^3+41607 s}{33554432 (-\beta )^{5/2}}+...
\end{align} 
with $s=2\n$ and the Floquet exponent $\n$ is now linked with the $a_D$-cycle. The WKB quantization condition translates into $s=2n+1$ where $n$ is the overtone. If we plug the dictionary in the previous expression, the latter becomes an equation for $\omega$ which can be numerically solved in order to find QNM frequencies. See second column in Table \ref{tabBHQNM}.

\subsection{Leaver's numerical method}
In order to confirm the results obtained with WKB approximation and with qSW, one can solve the differential equation exploiting the method of continuous fraction formulated by Leaver \cite{Leaver1, Leaver2}.

The radial equation \eqref{seq} can be transformed in such a way as to show two regular and one irregular singular point by setting\footnote{The two choices of sign cover two different allowed ranges in $r$.}:
\be
r=L\sqrt[4]{{\tilde{\omega}^2\over\tilde{\omega}^2-{\mu}^2}}\left(1+2y\pm2\sqrt{y(1+y)}\right)
\ee
leading to
\be\label{D1D521}
y(1+y)\Psi''(y)+{1+2y\over2}\Psi'(y)+\Big[2\b-\a+16\b y(1+y)\Big]\Psi(y)=0
\ee
that can be put in the standard form of a CHE for  $W(z)=e^{\pm 4 i \sqrt{\beta} z} \Psi(z)$ with $z=-y$.
The ansatz for the solution is:
\be
\Psi(y)=e^{i\omega y}y^{\sigma_+}(1+y)^{\sigma_-}\sum_{n=0}^\infty c_n\left({y\over1+y}\right)^n
\ee
The constants $\sigma_+$ and $\omega$ are determined by the behaviours near the horizon and at infinity respectively, while $\sigma_-$ is fixed by requiring that the recursion relation involves only three terms. We find:
\be
\omega^2=16 \b\quad,\quad\s_+=0,{1\over2}\quad,\quad \s_-=-\s_+-{1\over2}
\ee 
The two choices for $\s_+$ generate two different branches for QNMs, indeed we will see that $\s_+=0$ is associated to even overtones while $\s_+=1/2$  with odd ones. The recurrence relation is:
\begin{align}
&\a_n c_{n+1}+\b_n c_n+\g_n c_{n-1}=0\\\nn
&\a_n=(n+\s_++1)(n+\s_++1/2)\\\nn
&\b_n=-{1\over4}-2n^2-\a+2\b+n(-1+8i\sqrt{\b}-4\s_+)-\s_+(1+2\s_+)+2i\sqrt{\b}(1+4\s_+)\\\nn
&\g_n=(n+\s_+)(n+\s_+-1/2)
\end{align}
where we considered $c_{-1}=0$. The QNM frequencies associated to the overtone $n$ can be obtained by solving numerically the equation:
\be
\b_n={\a_{n-1}\g_n\over\b_{n-1}-{\a_{n-2}\g_{n-1}\over \b_{n-2}-...}}+{\a_n\g_{n+1}\over\b_{n+1}-{\a_{n+1}\g_{n+2}\over\b_{n+2}-...}}
\ee
Results coming from all the techniques that we exposed are collected in Table \ref{tabBHQNM}.
%


\begin{table}[h!]\small
\centering
\begin{tabular}{|c|c|c|c|c|}
\hline
    & WKB & Seiberg-Witten   &   Leaver  & Mathieu \\ \hline
$\ell=0$ &0.433013{-}0.25 I    & 0.498509{-}0.278493 I  &  0.498509{-}0.263183 I   & 0.501666{-}0.245563 I\\ \hline
$\ell=1$ & 0.968246{-}0.25 I  & 0.999627{-}0.255004 I   &  0.999643{-}0.253669 I  &  0.999883{-}0.253497 I \\ \hline
$\ell=2$ & 1.47902{-}0.25 I&   1.50114{-}0.250172 I  &  1.49988{-}0.251682 I   & 1.4999{-}0.251675 I \\ \hline
$\ell=3$  & 1.98431{-}0.25 I &  2.00112{-}0.249572 I   &  1.99994{-}0.250958 I  &  1.99995{-}0.250958 I\\ \hline
$\ell=4$ & 2.48747{-}0.25 I  &  2.50072{-}0.249845 I   & 2.49997{-}0.250617 I   &  2.49997{-}0.250617 I \\ \hline
$\ell=5$  & 2.98957{-}0.25 I &  2.98907{-}0.249864 I    &  2.99998{-}0.25043 I  & 2.99998{-}0.25043 I \\ \hline
$\ell=6$ & 3.49106{-}0.25 I &  3.50865{-}0.2499 I    &  3.49999{-}0.250317 I    & 3.49999{-}0.250317 I \\ \hline
$\ell=7$  & 3.99218{-}0.25 I  & 4.00383{-}0.249924 I   &   3.99999{-}0.250243 I    & 3.99999{-}0.250243 I\\ \hline
$\ell=8$  & 4.49305{-}0.25 I  & 4.50955{-}0.24994 I   &   4.49999{-}0.250192 I    &  4.49999{-}0.250192 I\\ \hline
$\ell=9$  &4.99375{-}0.25 I   &   5.00698{-}0.249951 I   &  5.{-}0.250156 I   &  5.{-}0.250156 I \\ \hline
$\ell=10$  & 5.49432{-}0.25 I  &  5.49877{-}0.24996 I  &  5.5{-}0.250129 I    & 5.5{-}0.250129 I\\ \hline
\end{tabular}
\caption{Prompt Ringdown modes of $D1-D5$ small BH for $n=0$, $M=0$ and $L=1$}\label{tabBHQNM}
\label{Table1}
\end{table}

\subsection{Absorption probability and cross section}

Using the connection formulae for Mathieu equation or equivalently for the DRDCHE one can determine the grey-body factor of small BHs as a function of $\omega$ and $\ell$. The analysis is very similar to the one for D3-branes \cite{Gubser:1998iu}.

%

Setting\footnote{The new $z$ is different from the one in Mathieu equation.}
 $$r={2 \sqrt{z} \over \sqrt{\tilde{\omega}^2-{\mu}^2}}$$
the wave equation can be written in the form of a DRDCHE
\be
W''(z) + \left({1\over z} - {p\over z^2} + {\varepsilon\over z^3}\right)W(z)=0
\ee
with $p=u-{1\over 4}$ and $\varepsilon= q = {\tilde\omega^4 L^4\over 16}
\left(1- {\mu^2\over \tilde\omega^2}\right)$

The DRDCHE has two irregular singularities: one at $z= 0$, corresponding to the `horizon' at $r=0$,  and the other one at $z\rightarrow \infty$, corresponding to asymptotically flat space-time at $r=\infty$.

The connection formula for the wave ingoing into the horizon at $z=0$ to the waves at $z=\infty$ reads \cite{Bonelli:2022ten}
$$
z^{3\over 4} e^{2i \sqrt{\varepsilon \over z}} W^{DRDC}_0(p, \varepsilon; \sqrt{z}) ={z^{1\over 4} \over 2\pi i} \sum_{\sigma=\pm} {\Gamma(1-2\sigma a(q))^2  \over -2\sigma a(q)} \varepsilon^{{1\over 4} + \sigma a(q)}
e^{-{\sigma\over 2} \partial_{a}{\cal F}} \times $$
$$  
\left[e^{2i \sqrt{z}}W^{DRDC}_0\left(p, \varepsilon; \sqrt{\varepsilon 
\over z}\right) + i e^{-2\pi i\sigma a(q)}  e^{-2i \sqrt{z}}W^{DRDC}_0\left(p, e^{2\pi i}\varepsilon; -\sqrt{\varepsilon 
\over z}\right)\right] 
$$
where 
\be 
W^{DRDC}_0(p, \varepsilon; x) = 1 +i {3+16p \over 16\sqrt{\varepsilon}} \sqrt{z} + {\cal O}(z)  
\ee 
is the DRDC Heun function, ${\cal F}={\cal F}(a,q, m_f; \hbar)$ is the NS pre-potential and $a(q)$ the fundamental (quantum-deformed) period. The wave at infinity can be written as
\be
\Psi_\infty \approx B(\omega) e^{- i \sqrt{\tilde\omega^2 - \mu^2} r} + C(\omega) e^{+ i \sqrt{\tilde\omega^2 - \mu^2} r}
\ee
with (using the connection formulae)
\be
B(\omega) = A \sin(\pi a_D - 2\pi a) \quad , \quad C(\omega) = A \sin(\pi a_D)
\ee
so that the absorption probability is given by  
\be
P(\omega) = 1 - {|\sin(\pi a_D)|^2 \over |\sin(\pi a_D - 2\pi a)|^2} 
\ee
Since for $\omega$ real $a$ (half of the Floquet exponent $\nu$) is real and $a_D=i\tau$ is purely imaginary, the expression simplifies to 
\be
P_{abs}(\omega) = 1 - {\sinh^2(\pi \tau) \over \sinh^2(\pi \tau) + \sin^2(2\pi a)}  = 
{\sin^2(2\pi a) \over \sinh^2(\pi \tau) + \sin^2(2\pi a)}
\ee
that clearly satisfies $0\le P_{abs}(\omega) \le 1$. The computation of $a$ or directly $\eta = \exp{2\pi i a}$, as in \cite{Gubser:1998iu}  is (almost) straightforward in terms of residues of the quantum SW differential or using {\it cum grano salis} the continuous fractions formula for the `energy' eigenvalue of Mathieu equation, summarised in \eqref{ufroma}. 

The computation of $a_D$ or $\chi = \exp{\pi i a_D}$, as in \cite{Gubser:1998iu}, is more laborious. Following \cite{Gubser:1998iu} one can write
\be
\chi = \exp{\pi i a_D}   =q^{-a} {\Gamma(a+\sqrt{u}+1) \Gamma(a-\sqrt{u}+1) \over \Gamma(-a+\sqrt{u}+1) \Gamma(-a-\sqrt{u}+1)} {v(-a)\over v(+a)}
\ee
where 
\be
v(+a) = \sum_{n=0}^\infty A_n(a)q^n
\ee
In turn, the coefficients are given by nested sums \footnote{To avoid confusion with other $a$'s, we call $B_p$ what was called $a_p$ in \cite{Gubser:1998iu} and denote their $r$ by $\sqrt{u}$.} 
\be
A_n(a) = \sum_{p_1=0}^\infty\sum_{p_2=2}^\infty ... \sum_{p_n=2}^\infty  B_{p_1+a} B_{p_1+p_2+a} ... B_{\Sigma_i p_i+a}
\ee 
where 
\be
B_{a} = {1\over [(a+1)^2-u][(a+2)^2-u]}
\ee 
that can be computed recursively. Starting from $A_0=1$ one finds
\be
A_1(a) = S[1](a) \quad , \quad A_2(a) = {1\over 2} S^2[1](a)-{1\over 2} S[2](a) - S[1,1](a) ...
\ee
where the `loop variables' are defined as
\be
S[k_0,k_1,...k_n](a) = \sum_{p=-\infty}^{+\infty} \prod_{j=0}^n B^{k_j}_{p+j}(a)
\ee
with the understanding that $B_{p}(a)=0$ for $p<0$, so that 
\begin{align}
S[1]&={2z+3\over (4{u}-1)[(1+z)^2-{u}]}+{\psi(z-\sqrt{u}+1)-\psi(z+\sqrt{u}+1)\over \sqrt{u}(4{u}-1)}\\\nn
S[2]&={35-70{u}+8{u}^2+84z-100{u}z+16{u}^2z+70z^2-56{u}z^2+20z^3-16{u}z^3\over(4{u}-1)^3[{u}-(z+1)^2]^2}+\\\nn
&+{(20{u}{-}1)(\psi(1{-}\sqrt{u}{+}z)-\psi(1{+}\sqrt{u}{+}z))\over2\sqrt{u}^3(4{u}{-}1)^3}+{(4{u}{+}1)(\psi^{(1)}(1{-}\sqrt{u}{+}z)+\psi^{(1)}(1{+}\sqrt{u}{+}z))\over2{u}(4{u}{-}1)^2}\\\nn
S[1,1]&={(10{u}-1)(\psi(1-\sqrt{u}+z)-\psi(1+\sqrt{u}+z))\over4\sqrt{u}^3({u}-1)(4{u}-1)^2}{-}{\psi^{(1)}(2-\sqrt{u}+z)+\psi^{(1)}(2+\sqrt{u}+z)\over 4{u}(4{u}-1)}+\\\nn
&-{4u^3(3 + 2z) + u^2(35-26z-36z^2-8z^3)-u (109 + 143z + 65z^2+ 10z^3) + 2(2 + z)^2 \over 4({u}-1)(4{u}-1)^2[{u}-(z+2)^2][{u}-(z+1)^2]^2}
\end{align}
Setting $\mu=0$ for simplicity and plugging in the expressions of $a$ and $a_D$ in terms of $q={\omega^4L^4\over 16}$ and $u = {1\over 4} (\ell + 1)^2- {1\over 2} \omega^2L^2$ one finds an expansion of the form
\be
P_\ell(\omega)  ={4\pi^2 \delta_\ell q^{\ell +1} \over (\ell!)^4(\ell+1)^2} \sum_{n=0}^\infty \sum_{k=0}^n P^{(\ell)}_{n,k}q^{n/2}\log^k(\hat{q})
\ee
with $ \hat{q} = q e^{4\gamma_E}  $ with $\gamma_{E}\approx 0.577216 =-\psi(1)$ the Euler-Mascheroni constant and $\delta_\ell=1$ for $\ell\neq 0$, while $\delta_0=256/225$ due to a subtle cancellation between powers of $\sqrt{q}$ in $a_D$ or rather in $2i \sin\pi a_D = \chi - \chi^{-1}$. 

With this proviso, the coefficient $P_{0,0}^{(\ell)}=1$ for any $\ell$ the first few other coefficients are:
\begin{align}
&P^{(0)}_{1,0}= {499\over 60}
\quad,\quad 
P^{(0)}_{1,1}= -{68\over 15}
\\\nn
&P^{(0)}_{2,0}={115477\over 1600}-{2224\pi^2\over 225} 
\quad, \quad 
P^{(0)}_{2,1}= -{3604\over 75} 
\quad,\quad 
P^{(0)}_{2,2}= {856\over 75} 
\\\nn
&P^{(0)}_{3,0}={230199619-59928064\pi^2-33945600\z(3)\over 432000} \quad, \quad 
P^{(0)}_{3,1}=-{7(3595071- 504832\pi^2)\over54000}
\\\nn 
&P^{(0)}_{3,2}= {170222\over 1125}
\quad,\quad  
P^{(0)}_{3,3}=-{75616\over 3375} \\\nn
\end{align}
\begin{align}
&P^{(1)}_{1,0}= {31\over3} 
\quad,\quad 
P^{(1)}_{1,1}= -2 \\\nn
&P^{(1)}_{2,0}={4763\over72}-{4\pi^2\over 3} 
\quad, \quad 
P^{(1)}_{2,1}= -{67\over3}
\quad,\quad 
P^{(1)}_{2,2}=2 \\\nn
&P^{(1)}_{3,0}={17399\over 48}-16\pi^2-{32\z(3)\over 3}
\quad,\quad 
P^{(1)}_{3,1}=-{1379\over9}+{8\pi^2\over 3}
\quad,\quad P^{(1)}_{3,2}=24
\quad,\quad P^{(1)}_{3,3}=-{4\over3}\\\nn
\end{align}
\begin{align}
&P^{(2)}_{1,0}= {649\over72} \quad,\quad P^{(2)}_{1,1}= -{4\over3} \\\nn
&P^{(2)}_{2,0}={985567\over20736}-{16\pi^2\over27} \quad, \quad 
P^{(2)}_{2,1}= -{337\over27}\quad,\quad P^{(2)}_{2,2}={8\over9} \\\nn
&P^{(2)}_{3,0}={22584863\over116640}-{466\pi^2\over81}-{256\z(3)\over81}
\quad,\quad P^{(2)}_{3,1}=-{15843885-184329\pi^2\over233280}\quad,\quad P^{(2)}_{3,2}={233\over 27}
\\\nn
& P^{(2)}_{3,3}=-{32\over81}\\\nn
\end{align}
Note that the coefficients do not enjoy uniform transcendentally, as for D3-branes \cite{Gubser:1998iu}. Note also that the leading dependence on $\omega$ comes from $q^{\ell+1} \approx \omega^{4\ell+4}$.

Using a generalisation of optical theorem to $d=D-1$ non-compact transverse spatial directions, see \cite{Gubser:1998iu,Gubser:1997qr}, that relies on partial wave expansion in terms of the relevant spherical harmonics (Gegenbauer polynomials, related to characters of $SU(2)$ \cite{Bianchi:2017sds}) and Bessel functions one has (for massless probes $k=\omega$)
\be
\sigma_\ell(\omega) =  {2^{d-2}\pi^{{d\over 2}-1}  \over 3 \omega^5} (\ell+{d\over 2}-1)\Gamma({d\over 2}-1)\left(^{\ell+d-3}_{\quad\ell\quad}\right) (1-|S_\ell|^2)
\ee

Replacing $1-|S_\ell|^2=P_\ell$ and putting $d=4$ for for D3/D3' or D1/D5 $d=4$ one finally finds
\be
\sigma_\ell(\omega) =  {4\pi  \over \omega^3} (\ell+1)^2 P_\ell(\omega) 
\ee
with the absorption probably computed above. It is worth recalling how crucial this kind of analysis was for substantiating the AdS/CFT correspondence in its early days. Yet the role of the Couch-Torrence symmetry \cite{CouchTorr} $z\leftrightarrow -z$ or $r\leftrightarrow L^2/r$ that exchanges the horizon with flat infinity, keeping the photon-sphere fixed, has not been fully exploited in this context \cite{Bianchi:2021yqs, Bianchi:2022wku}.

\subsection{Tidal Love numbers}
Another interesting observable  is tidal deformability. For static waves $\omega=0$ and  in a spherically symmetric case, such as a small BH, the tidal Love number \cite{Love1909.0008, Bonelli:2021uvf, Cvetic:2021vxa, Pereniguez:2021xcj, Consoli:2022eey} is determined by the ratio of the term decaying at infinity (`response') wrt the one growing at infinity (`source'). 

For $\omega=0$ and $\mu=0$ the two behaviours are 
\be 
\Psi(r)\approx r^{\ell + {3\over 2}} \quad {\rm and} \quad \Psi\approx r^{-\ell - {1\over 2}} 
\ee

In fact the exact solution is 
\be 
\Psi(r) = A_\ell r^{\ell + {3\over 2}} + B_\ell r^{-\ell - {1\over 2}} 
\ee
where $A_\ell$ plays the role of the `source' and $B_\ell$ of the `response'. The static Love number is ${\cal L}_\ell = B_\ell/A_\ell$. Imposing regularity at the horizon ($r=0$) implies $B_\ell=0$ and as a consequence 
\be
{\cal L}^{smallBH}_{\ell, stat} =0
\ee 
Similarly to Kerr-Newman BHs in 4-d and extremal BHs in 5-d \cite{Pereniguez:2021xcj}. 

In the dynamical case $\omega\neq0$ the definition of tidal Love numbers is somewhat ambiguous, since the wave behaves as $e^{\pm i \omega r}$ at infinity. 
In the (quasi)static limit $\omega r <<1$ and $M/r<<1$\footnote{$M$ represents the mass/length scale of the compact object.} ($M<<r<<1/\omega$) the wave behaves according to
$$
\Psi(r) \approx r^{\lambda_-} {\cal R}(\omega r) + r^{\lambda_+} {\cal S}(\omega r)  
$$
where ${\lambda_\pm}$ depend on $\omega$ and $\ell$. 
One can generalise the notion of Love number and define a Love `function' \cite{Consoli:2022eey}
$$
{\cal L}(\omega r) = {{\cal R}(\omega r) \over {\cal S}( \omega r)}  
$$ 
that can be derived using the by-now available connection formulae. In the case of $N_f=(1,2)$ q-SW curves, setting $\omega =0$ one finds \cite{Consoli:2022eey}
\be\label{lovefunc}
{\cal L}(0)={\cal L}_0 = L+i \Lambda = {\Gamma(-2 a) \Gamma(1+m_1+a) \Gamma(1+m_2+a) \over 
\Gamma(2 a) \Gamma(1+m_1-a) \Gamma(1+m_2-a)}
\ee
where the imaginary part $\Lambda$ account for dissipative effects.


In order to put \eqref{D1D521} in canonical $N_f=(1,2)$ q-SW form, we have to perform several steps.

First we have to redefine $\Psi(y)\rightarrow y^{-1/4}(1+y)^{-1/4} \Psi(y)$, so that
\be
\Psi''(y)+{256\b y^4+512\b y^3+4(1-4\a+72\b)y^2+4(1-4\a+8\b)y+3\over16y^2(1+y)^2}\Psi(y)=0
\ee

Furthermore if we map $y=-1/z$ and expand in partial fractions we obtain:
\begin{align}
\Psi''(z)&+\left({16\b\over z^4}+{1-4\a+8\b\over 4z^2}+{3\over 16(z-1)^2}+{1-4\a+8\b\over 4z}+{-1+4\a-8\b\over4(z-1)}\right)\Psi(z)=0
\end{align}
which corresponds to  $Q_{2,1}(z)$ in \cite{Consoli:2022eey} that can be written in terms of the CFT data (see Appendix \ref{AGT})  as
\begin{align}
\Psi''(z)&+\left({-}{q^2\over4z^4}{+}{q  c\over z^3}{+}{1-4u-2q+4k_0 q\over 4z^2}{+}{\d_{k_0}\over(z-1)^2}+\right.\\\nn
&\left.+{1-4u-2q+4k_0 q+4\d_{k_0}-4\d_{p_0}\over4z}+{-1+4u+2q-4k_0 q-4\d_{k_0}+4\d_{p_0}\over4(z-1)}\right)\Psi(z)=0
\end{align}

For D1-D5 small BH  the dictionary is thus:
\be\label{dict2}
q{=}{-}8 i\sqrt{\b}\quad,\quad c=0\quad,\quad u=\a+2i\sqrt{\b}-2\b\quad,\quad \d_{k_0}{=}\d_{p_0}{=}{3\over16}
\ee
which implies $m_1 =-1/2$, $m_2=0$ and $m_4=1/2$, while $m_3$ is undetermined and can be decoupled (taken to infinity).
Mapping to $N_f=(1,2)$ q-SW, the $a$-cycle can be written as:
\be\label{acycle21}
a=\sqrt{u}-{1\over 8\sqrt{u}}q-{1+2u\over128u^{3/2}}q^2-{1+2u\over1024u^{5/2}}q^3+{5+7u-8u^2-6u^3\over32768(u-1)u^{7/2}}q^4+...\footnote{The dictionary (1,2) SW-BH is slightly different: q is the same as \eqref{dict2}, $u=\a-2i\sqrt{\b}-2\b$, $m_1=1/2$, $m_2=m_3=0$}
\ee
In order to obtain the Love function, we have to plug in \eqref{lovefunc} the dictionary \eqref{dict1}, \eqref{dict2} and \eqref{acycle21} together with \eqref{mathieudict}.

In the static limit, using the duplication formula for $\Gamma$ functions, we obtain:
\be
\mathcal{L}_0= {\Gamma(-2a) \Gamma({1\over2}+ a) \Gamma(1+a)\over \Gamma(2a) \Gamma({1\over2}-a) \Gamma(1-a)} = - 2^{-4a} = - 2^{-4\sqrt{(\ell+1)^2+L^2\m^2}} 
\ee
 Notice that a non-zero result is in disagreement with the `rigorously' static case. As observed in \cite{Consoli:2022eey} the two limits do not commute. Computing the `dynamical' Love number with $\omega \neq 0$ via the `Love function'  \cite{Consoli:2022eey} (defined for $L<<r<<1/\omega$) by imposing regularity at the horizon and then taking the ratio of the coefficient of the term decaying at infinity (as $r^{-\ell - {1\over 2} - \Delta\ell (\omega)}$) and of the term growing at infinity (as $r^{+\ell + {3\over 2} + \Delta\ell(\omega)}$) yields the above non-zero result. The mathematical reason is that the singularity structure of the equation changes though in the `intermediate' region $L<<r<<1/\omega$ one can still keep only the $1/r^2$ term in the potential. Physics suggests that the relevant Love number be the dynamical one since a `static' perturbation with $\omega=0$ exactly is hard to conceive in an astrophysical setting.

Since the small BH has no ergo-region we do not expect super-radiance for neutral probes. Yet one can study charge super-radiance \cite{Brito:2015oca,Aalsma:2022knj}. This is beyond the scope of the present investigation.

\section{Probing and re-probing circular fuzz-ball with particles and waves}

We are ready to discuss the main topic of the present investigation: scalar perturbations of a fuzz-ball in $D=5$ with circular profile of radius $a_f$. 

The metric of the D1/D5 circular fuzz-ball (in the democratic D3/D3' frame) reads \cite{Lunin:2001jy}:
\begin{align}
ds^2=&H^{-1}\Big[-(dt+\omega_\phi d\phi)^2+(dz+\omega_\psi d\psi)^2\Big]+H\Big[(\r^2+a_f^2\cos^2\q)\left({d\r^2\over\r^2+a_f^2}+d\q^2\right)+\\\nn
&+\r^2 \cos^2\q d\psi^2+(\r^2+a_f^2)\sin^2\q d\phi^2\Big] + \left({H_1\over H_5}\right)^{1\over2}dz^2_{1,2}+\left({H_5\over H_1}\right)^{1\over2}dz^2_{3,4}
\end{align}
where
\begin{align}
H=\sqrt{H_1 H_5}\quad,\quad H_i=1+{L_i^2\over\r^2+a_f^2\cos^2\q}\\\nn
 \omega_\phi={a_fL_1L_5\sin^2\q\over\r^2+a_f^2\cos^2\q}\quad,\quad\omega_\psi={a_f L_1 L_5\cos^2\q\over \r^2+a_f^2\cos^2\q}
\end{align}
and the spheroidal coordinates are related to the standard cartesian coordinates by
\be
x_1+ix_2 = \rho \cos\theta e^{i\psi} \quad , \quad x_3+ix_4 = \sqrt{\rho^2+a_f^2} \sin\theta e^{i\phi}
\ee
so that 
\be
\rho \cos\theta = r \cos\tilde\theta \quad , \quad \sqrt{\rho^2+a_f^2} \sin\theta = r \sin\tilde\theta
\ee
with $r$ and $\tilde\theta$ the usual radial and polar angular coordinates.

It is easy to check that in the limit $a_f\rightarrow 0$ one gets back the small BH metric, while for $a_f\neq 0$ the solution is non-singular and horizon-less \cite{Lunin:2001jy}. Notice however that, contrary to the small BH, the circular fuzz-ball carries angular momentum. This will play a role in the following. 

The generalized momenta are:
\be
P_t=-{\dot{t}+\omega_\phi \dot{\phi}\over H}=-E\quad,\quad P_z={\dot{z}+\omega_\psi \dot{\psi}\over H}\quad,\quad P_\rho={H(\rho^2+a_f^2\cos^2\q)\over\rho^2+a_f^2}\dot{\rho}
\ee
$$P_\q=H(\rho^2+a_f^2\cos^2\q)\dot{\q}\quad,\quad P_\phi=-{\omega_\phi\over H}\dot{t}+\Bigg[H(\rho^2+a_f^2)\sin^2\q-{\omega_\phi^2\over H}\Bigg]\dot{\phi}=J_\phi$$
$$P_\psi={\omega_\psi\over H}\dot{z}+\left(H\rho^2\cos^2\q+{\omega_\psi^2\over H}\right)\dot{\psi}=J_\psi\quad,\quad P_{1,2}=\left({H_1\over H_5}\right)^{1\over2}\dot{z}_{1,2}\quad,\quad P_{3,4}=\left({H_5\over H_1}\right)^{1\over2}\dot{z}_{3,4}$$
where the only non conserved `dynamical' momenta are $P_\rho$ and $P_\q$. If we invert the previous relations, we obtain:
\be
\dot{t}=H E-{\omega_\phi(J_\phi +E\omega_\phi)\over H(\rho^2+a_f^2)\sin^2\q}\quad,\quad \dot{\phi}={J_\phi+E \omega_\phi\over H(\rho^2+a_f^2)\sin^2\q}
\ee
$$\dot{z}=H P_z-{\omega_\psi(J_\psi-P_z\omega_\psi)\over H\rho^2\cos^2\q}\quad,\quad \dot{\psi}={J_\psi-P_z \omega_\psi\over H\rho^2\cos^2\q}$$
$$\dot{\rho}={\rho^2+a_f^2\over H(\rho^2+a_f^2\cos^2\q)}P_\r\quad,\quad \dot{\q}={1\over H(\r^2+a_f^2\cos^2\q)}P_\q$$
$$\dot{z}_{1,2}=\left({H_5\over H_1}\right)^{1\over2}P_{1,2}\quad,\quad \dot{z}_{3,4}=\left({H_1\over H_5}\right)^{1\over2} P_{3,4}$$

\subsection{Massless Geodesics}
For zero KK momenta along the internal torus $P_I=0$, massless geodesics in Hamiltonian form are determined by
\be
P_\r^2(\r^2+a_f^2)+{\mathcal{L}_\psi^2\over \r^2}-{\mathcal{L}_\phi^2\over\r^2+a_f^2}-\mathcal{E}^2 (\r^2+a_f^2+L_1^2+L_5^2)+P_\q^2+{J_\psi^2\over\cos^2\q}+{J_\phi^2\over\sin^2\q}+\mathcal{E}^2a_f^2\sin^2\q=0
\ee
where 
\be
\mathcal{E}^2=E^2-P_z^2 \quad,\quad \mathcal{L}_\phi^2=(J_\phi a_f - E L_1L_5)^2\quad,\quad \mathcal{L}_\psi^2=(J_\psi a_f - P_z L_1L_5)^2 
\ee
The conjugate momenta $E=-P_t$, $P_z$, $J_\phi= P_\phi$, and $J_\psi=P_\psi$ are conserved but the spherical symmetry $SO(3)\times SO(3)$ of the small BH is broken to $SO(2)_\phi \times SO(2)_\psi$ by the very presence of the circular profile.   
Despite the reduced isometry, very much as for Kerr BHs, separation is possible \`a la Carter \cite{Chervonyi:2013eja, Bianchi:2017sds}:
\begin{align}
&P_\r^2=Q_R(\r)=\mathcal{E}^2\left(1+{L_1^2+L_5^2\over\r^2+a_f^2}\right)+{\mathcal{L}_\phi^2\over(\r^2+a_f^2)^2}-{\mathcal{L}_\psi^2\over \r^2(\r^2+a_f^2)}-{K^2\over \r^2+a_f^2}\\\nn
&P_\q^2=Q_A(\q)=K^2-{J_\psi^2\over \cos^2\q}-{J_\phi^2\over\sin^2\q}-\mathcal{E}^2a_f^2\sin^2\q
\end{align}
Notice that with this choice, the separation constant $K^2$ is positive\footnote{This is not so for other `standard' choices e.g. $A= K^2 - a_f^2{\cal E}^2$.} for real $\mathcal{E}$ and plays the role of the total angular momentum including frame-dragging effects. Notwithstanding the separation of the radial dynamic from the polar angular one, geodesics are non-planar in general. In fact they typically give rise to quite non-trivial motion in any direction. Some simplification arises when the motion takes place in a given `hyperplane' with $\theta = \theta_0$. We will call these geodesics `shear-free', borrowing the terminology from the familiar case of Kerr BHs \cite{ChandraBH}.
We will start with the two simplest cases: $\theta=0$ and $\theta=\pi/2$.

\subsubsection{$\q=0$ and $J_\phi=0$}
\begin{figure}[h]
\begin{minipage}[b]{7cm}
\centering
\includegraphics[width=8.5cm]{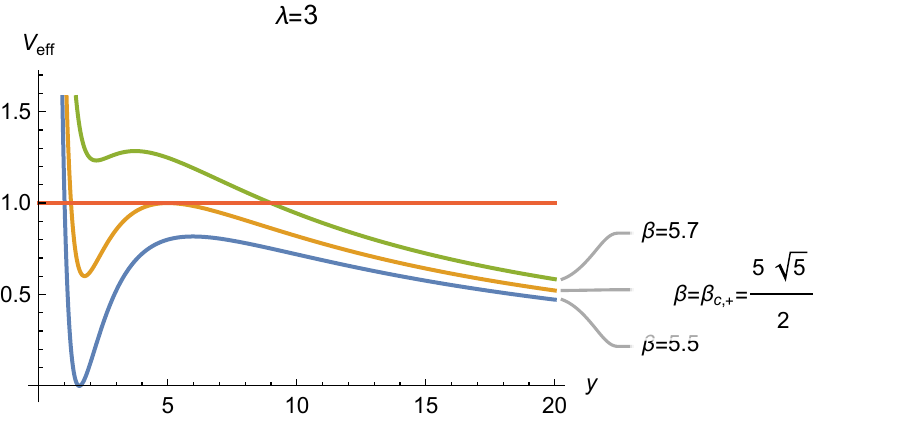}
\end{minipage}
\ \hspace{2mm} \hspace{3mm} 
\begin{minipage}[b]{7cm}
\centering
\includegraphics[width=8.5cm]{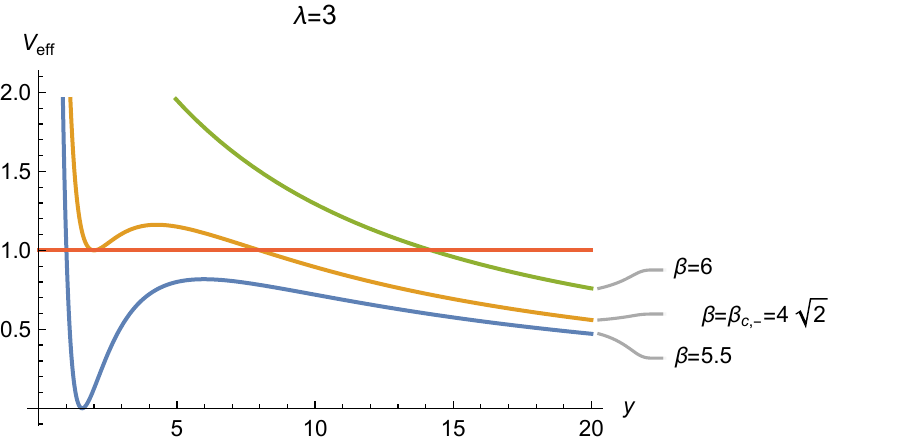}
\end{minipage}
\caption{Left pannel: plot of the effective potential for $\lambda=3$ which implies $y_{c,+}=5$ and $\b_{c,+}=5\sqrt{5}/2$. Yellow line correspond to $\beta=\b_{c,+}$, green line to $\b=5.7>\b_{c,+}$ and blue line to $\b=5.5<\b_{c,+}$.
Right pannel: for $\lambda=3$ the minimum is located at $y_{c,-}=2$ and $\b_{c,-}=4\sqrt{2}$. Yellow line correspond to $\beta=\b_{c,-}$, green line to $\b=6>\b_{c,-}$ and blue line to $\b=5.5<\b_{c,-}$}\label{pot0eqplane}
\end{figure}

For simplicity but without significant loss of generality, we set $P_z=0$ and $L_1=L_5=L$. In the hyper-plane $\q=0$, orthogonal to the plane of the circle, the angular equation forces one to take $J_\phi=0$, so that $K^2=J_\psi^2$. The potential reads:
\be
Q_R(y)=E^2\Bigg[1-\left({\b^2\over y}-{2\l^2\over 1+y}-{\l^4\over(1+y)^2}\right)\Bigg]=E^2\Big[1-\mathcal{V}_{eff}(y)\Big]
\ee
where we defined
\be\label{notation}
y={\r^2\over a_f^2}\quad,\quad \b={K\over a_f E} = {b\over a_f} = {|J_\psi|\over a_f E} \quad,\quad \l={L\over a_f}
\ee
Since 
\be
\mathcal{V}_{eff}(y)\sim {\beta^2-2\l^2\over y}+\mathcal{O}\left(y^{-2}\right)\quad\text{as}\quad y\rightarrow\infty
\ee
in order to have a photon-sphere one must consider $\beta^2>2\l^2$. Critical geodesics are related to  double roots of $Q_R$  
\be
\mathcal{V}_{eff}(y)=1 \quad , \quad \mathcal{V}'_{eff}(y)=0
\ee
considered as a system in $y_c$ and $\b_c$. The solutions are:
\be
y_{c,+}={1\over2}(\l^2-2+\l\sqrt{\l^2-8})\quad,\quad \b_{c,+}^2={8(11\l-14\l^3+2\l^5+\sqrt{\l^2-8}(1-6\l^2+2\l^4)\over(\l+\sqrt{\l^2-8})^3}
\ee
and
\be
y_{c,-}={1\over2}(\l^2-2-\l\sqrt{\l^2-8})\quad,\quad \b_{c,-}^2={8(11\l-14\l^3+2\l^5-\sqrt{\l^2-8}(1-6\l^2+2\l^4)\over(\l-\sqrt{\l^2-8})^3}
\ee
for $\lambda>2\sqrt{2}$ we have in general $y_{c,+}>y_{c,-}$ which are respectively a maximum and a minumum (see Fig. \ref{pot0eqplane}).

%
Notice that, when both are present, the maximum at $y=y_{c,+}$ corresponds to the photon-sphere (light-ring) comprising unstable circular orbits, while the minimum at $y=y_{c,-}$ to a internal stable circular orbit. 


\subsubsection{$\q=\pi/2$ and $J_\psi=0$}
If we consider geodesics in the hyper-plane $\q=\pi/2$, where the circular profile is located, we are forced to set $J_\psi=0$, so the angular equation reduces to: 
\be
\b^2=\b_\phi^2+1\quad,\quad \b_\psi={J_\psi\over a_f E}=0
\ee
Using the notation introduced in \eqref{notation}, the radial function reads
\be
Q_R=E^2\Bigg[1-\left({\b_\phi^2+1-2\l^2\over(1+y)}-{(\l^2-\b_\phi)^2\over(1+y)^2}\right)\Bigg]=E^2\Big[1-\mathcal{V}_{eff}(y)\Big]
\ee
The presence of the photon-sphere requires $\b_\phi^2>2\l^2-1$ {\it i.e.} $\b_\phi>+\sqrt{2\l^2-1}$ (co-rotating) or $\b_\phi<-\sqrt{2\l^2-1}$ (counter-rotating). The critical regime is encoded in the system $Q_R=Q'_R=0$, whose solutions are:
\be\label{critcond}
\beta^{corot}_{\phi, c} = 2\lambda -1\quad,\quad y^{corot}_c=\lambda(\lambda-2)
\ee
$$
\beta^{counter}_{\phi, c} = -2\lambda - 1 \quad,\quad y^{counter}_c=\lambda(\lambda+2)
$$
$$
\beta_{\phi, c} = 1 \quad y_c=-\lambda^2 
$$

\begin{figure}[h]
\begin{minipage}[b]{7cm}
\centering
\includegraphics[width=8.5cm]{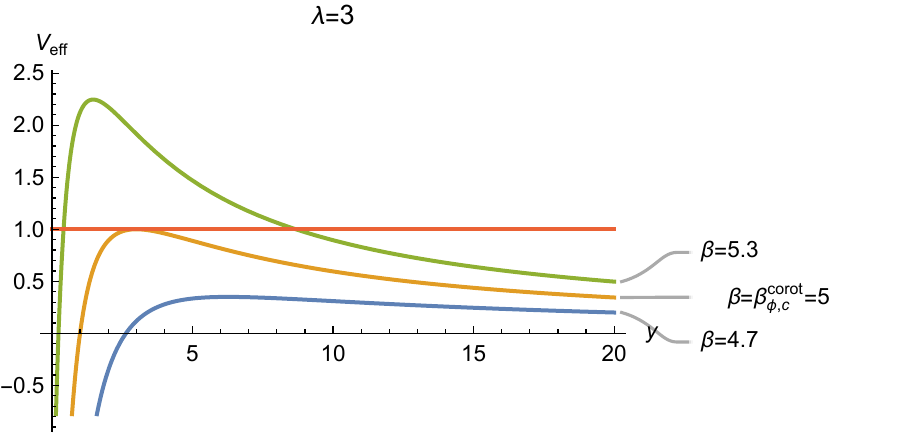}
\end{minipage}
\ \hspace{2mm} \hspace{3mm} 
\begin{minipage}[b]{7cm}
\centering
\includegraphics[width=8.5cm]{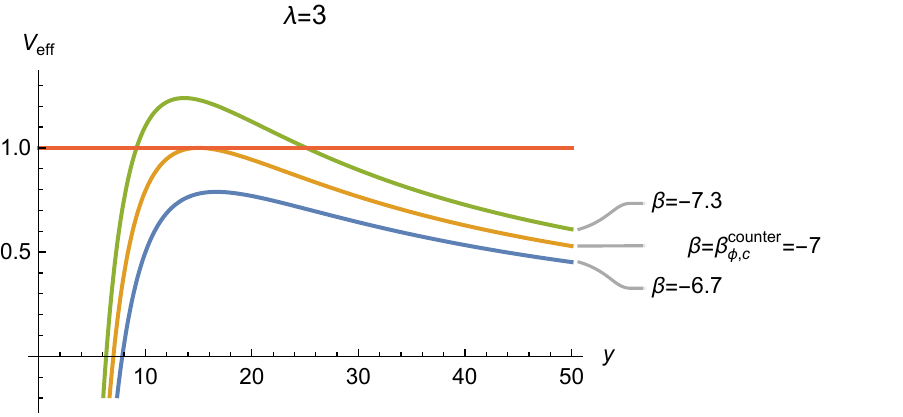}
\end{minipage}
\caption{Left pannel: plot of the effective potential for $\lambda=3$ which implies $y_c^{corot}=3$ and $\b_c^{corot}=5$. Yellow line corresponds to $\beta=\b_c^{corot}$, green line to $\b=5.3>\b_c^{corot}$ and blue line to $\b=4.7<\b_c^{corot}$.
Right pannel: for $\lambda=3$ we have $y_{c}^{counter}=15$ and $\b_{c}^{counter}=-7$. Yellow line corresponds to $\b_{c}^{counter}$, green line to $\b=-7.3<\b_{c}^{counter}$ and blue line to $\b=-6.7<\b_{c}^{counter}$}\label{potpimezzieqplane}
\end{figure}

In addition we must require $y_c>0$ that excludes the last case and imposes $\lambda\ge 2$ for the co-rotating solution. See Fig. \ref{potpimezzieqplane}.

\begin{figure}[h]
\centering
\includegraphics[width=8.5cm]{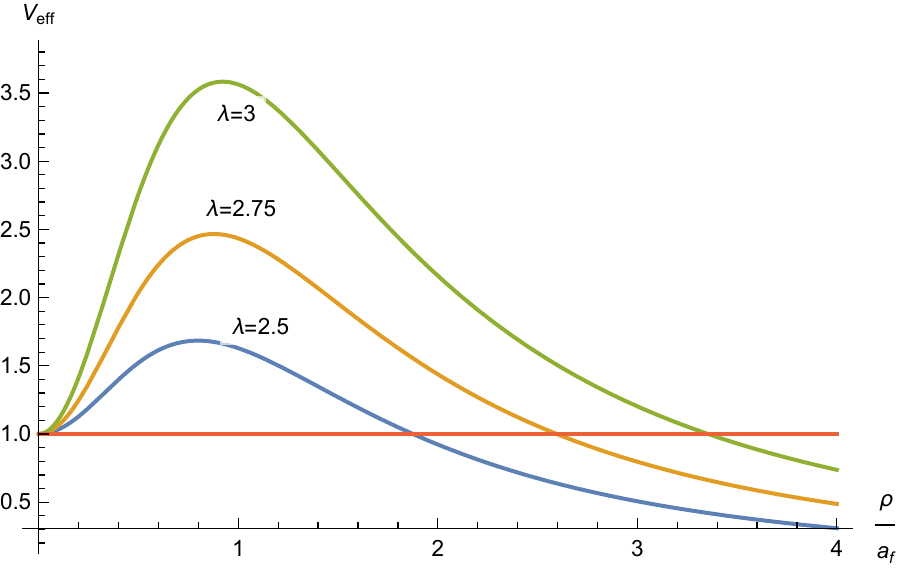}
\caption{Plot of the effective potential allowing the special geodesics at $\rho=0$ with $E= {2 a_f J_\phi  \over L^2+a_f^2}$ for different values of $\l$ }\label{potpimezzierho0}.
\end{figure}

The peculiar nature of the geometry in the $\theta =\pi/2$ plane, where space `ends' with a cap at $\rho = 0$, representing a circle of radius $a_f$ in that plane, allows for `critical' geodesics at $y = 0$. Indeed it is easy to check that setting $\rho = 0$ and $P_\rho=0$ one finds a solution for 
\be
E= {2 a_f J_\phi  \over L^2+a_f^2} =  a_f\dot\phi
\ee
Such kind of geodesics correspond to meta-stable modes as shown by the Fig. \ref{potpimezzierho0} for different values of $\lambda$.
\subsubsection{Radial fall and general geodesics}

Due to its angular momenta, the circular fuzz-ball drags particles even if they have $J_\phi=J_\psi=0$, since in this case $K^2 = P_\theta^2 + a_f^2 \omega^2\sin^2\theta \ge 0$ (strictly if $\theta \neq 0$).

For $\theta=0$ (constant) so that  $ P_\theta=0$ and $J_\phi=J_\psi=0$, $K^2=0$ one has `polar radial fall'. It is easy to check that in this case $\dot\rho =E$ `free fall' (in `proper' time). Since $\dot{t}=H E$ in this case,  and for $L_1=L_5=L$ one finds
\be
t - t_0 = \rho + {L^2\over a} \arctan{\rho\over a}
\ee

In general (for $L_1=L_5=L$)
\be
P_\rho = \dot\rho {\rho^2 + a_f^2\cos^2\theta + L^2 \over \rho^2 + a_f^2}
\ee
and   
\be
P_\theta = \dot\theta (\rho^2 + a_f^2\cos^2\theta + L^2 )
\ee
so much so that
\be
{d\rho\over (\rho^2 + a_f^2) P_\rho} = \pm  {d\theta\over P_\theta}
\ee
For $J_\phi=J_\psi=0$, $P_\theta=\sqrt{K^2- a_f^2E^2\sin^2\theta}$, while
$$
P^2_\rho = E^2 \left(1 + {L^2\over \rho^2 + a_f^2}\right)^2 - {K^2\over \rho^2 + a_f^2}
$$
Combining the two, one finds
\be
{d\rho\over \sqrt{E^2 (\rho^2 + a_f^2 +L^2)^2 - K^2(\rho^2 + a_f^2) }} = \pm  {d\theta\over \sqrt{K^2- a_f^2E^2\sin^2\theta}}
\ee
both can be reduced to incomplete elliptic integrals of the first kind. For the polar angle one finds
\be
\int^{\theta}_{\theta_0}{d\theta'\over K\sqrt{1- {a_f^2E^2\over K^2} \sin^2\theta'}}  = 
{1\over K} \Big[ F\left(\theta; {a_fE\over K}\right) - F\left(\theta_0; {a_fE\over K}\right)\Big]
\ee
while for the radial dynamics one has 
\be
\int^{\rho}_{\rho_0} {d\rho'\over 
\sqrt{E^2 \r^4{+}[2E^2(a_f^2{+}L^2){-}K^2]\r^2{+}E^2(a_f^2{+}L^2)^2{-}K^2a_f^2 } }=
\ee
$$={1\over E\r_+}\int_{\r_0\over \r_-}^{\r\over \r_-}{dx\over\sqrt{(1-x^2)\Big[1{-}\left({\r_-\over\r_+}\right)^2x^2\Big]}}={1\over E\r_+}\Big[F\left({\r\over\r_-};{\r_-\over\r_+}\right)-F\left({\r_0\over\r_-};{\r_-\over\r_+}\right)\Big]$$
with 
$$
\r^2_{\pm}=-a_f^2-L^2+{K^2\over2E^2}\left(1\pm\sqrt{1-4{E^2L^2\over K^2}}\right)
$$
\subsubsection{ Shear-free geodesics}
Shear-free geodesics \cite{ChandraBH} with $\theta=\theta_0$, are possible when $P_\theta =0$  {\it i.e.} 
\be
K^2_{s-free} =  {J_\phi^2 \over \sin^2\theta_0} + {J_\psi^2 \over \cos^2\theta_0} + a_f^2E^2\sin^2\theta_0
\ee
and $P_\theta' =0$ {\it i.e.} 
\be
0 =  - {J_\phi^2 \over \sin^3\theta_0} \cos\theta_0 + {J_\psi^2 \over \cos^3\theta_0} \sin\theta_0+ a_f^2E^2\sin\theta_0 \cos\theta_0
\ee

The radial momentum does not simplify much and leads to incomplete elliptic integrals of the first kind, as above. 

Barring the special cases $\theta_0 =0$ or $\pi/2$, that we have already analysed in detail, and setting $\xi=\cos^2\theta_0$, $\beta^2=K^2/a_f^2{\cal E}^2$, $\beta_\phi^2 = J_\phi^2/a_f^2{\cal E}^2$ and $\beta_\psi^2 = J_\psi^2/a_f^2{\cal E}^2$ one finds
\be
{\beta_\phi^2 \over (1-\xi)^2} - {\beta_\psi^2\over \xi^2} = 1
\ee
and 
\be
{\beta_\phi^2 \over 1-\xi} + {\beta_\psi^2 \over \xi} + 1 -\xi= \beta^2
\ee

Solving the first for $\xi = f(\beta_\phi, \beta_\psi)\in (0,1)$\footnote{This condition imposes restrictions on the allowed values of $ \beta_\phi, \beta_\psi$.} and plugging into the second yields the special values of $\beta_{s{-}f}$ in terms of  $\beta_\phi, \beta_\psi$ {\it i.e.} of  $K^2_{s{-}f}$ in terms of $J^2_\phi$, $J^2_\psi$ and $a_f^2{\cal E}^2$.
 
Otherwise, one can solve for $\beta_\phi$ and $\beta_\psi$ in terms of $\xi$ and $\beta$ and find
\begin{align}
\beta_\phi^2 = (1-\xi)^2 (\beta^2 + 2\xi -1)  \quad &\sim \quad J_\phi^2 = \sin^4\theta_0 [K^2 + a_f^2{\cal E}^2(2 \cos^2\theta_0 -1)] \\ \nn
\beta_\psi^2= \xi^2 (\beta^2 + 2\xi -2)  \quad &\sim \quad J_\psi^2 = \cos^4\theta_0 [K^2 - 2a_f^2{\cal E}^2\sin^2\theta_0] 
 \end{align}
For $K^2 >2a_f^2{\cal E}^2$ ($\beta>\sqrt{2}$) $\theta_0$ can take any value. For $K^2 < 2a_f^2{\cal E}^2$ ($\beta<\sqrt{2}$) $\theta_0$ is constrained by  $\theta_0< \arccos\left(1 - {K^2\over 2a_f^2{\cal E}^2}\right)$. 
 
 For small $a_f$, $|\beta_\phi|, |\beta_\psi|, \beta>>1$ and one finds
\be
(1-\xi)|\beta_\psi|\approx \pm \xi |\beta_\phi|\quad \sim \quad 
\xi \approx {|\beta_\psi| \over |\beta_\psi| \pm |\beta_\phi|}
\ee
(only the + sign is acceptable) and then
\be
|\beta_\phi| |\beta_\psi| +  \beta_\psi^2 - \beta^2 \xi \approx 0 \quad \sim \quad \beta^2 = \left[|\beta_\phi| |\beta_\psi| +  \beta_\psi^2 \right] {|\beta_\psi|+ | \beta_\phi|\over |\beta_\psi| } = [ |\beta_\psi| +  |\beta_\phi|]^2
\ee
or 
\be
\beta= |\beta_\phi|+ |\beta_\psi| \quad \sim \quad K = |J_\phi| +|J_\psi|
\ee

Clearly as soon as $a_f\neq 0$ the condition gets modified.

Shear-free geodesics will play a role in the study of wave perturbations with $\ell= |m_\phi|+|m_\psi|$ which is the minimal value of $\ell$ for given $m_\phi$ and $m_\psi$, since in  general $\ell= |m_\phi|+|m_\psi| + 2r$ with $r=0,1,...$.

\subsection{Wave equation}
Let us now pass and consider the Klein-Gordon equation for massless scalar waves in the D1-D5 fuzz-ball background. The wave equation can be separated as follows:
\begin{align}\label{eqfuzz}
&\Bigg\{{1\over \r}\partial_\r\Big[\r(\r^2+a_f^2)\partial_\r\Big]+\tilde{\omega}^2(\r^2+a_f^2)\left(1+{L_1^2+L_5^2\over\r^2+a_f^2}\right)+{\mathcal{L}_\phi^2\over\r^2+a_f^2}-{\mathcal{L}_\psi^2\over\r^2}-K^2\Bigg\}R(\r)=0\\\nn
&\Big[\nabla_{S_3}^2-\tilde{\omega}^2 a_f^2\sin^2\q+K^2\Big]S(\q)=0
\end{align}
where
\be
\nabla_{S_3}^2={1\over \sin(2\q)}\partial_\q\Big[\sin(2\q)\partial_\q\Big]-{m_\phi^2\over\sin^2\q}-{m_\psi^2\over \cos^2\q}
\ee
Let's transform the radial equation after introducing:
\be
{\r^2\over a_f^2}=y\quad,\quad {L_1^2+L_5^2\over a_f^2}=\lambda\quad,\quad {{\lambda}^2_{\phi,\psi}}={\mathcal{L}_{\phi,\psi}^2\over a_f^2}
\ee
so that the differential equation in canonical form becomes:
\be
\psi''(y)+\Bigg[{1{-}{{\lambda}^2_\psi}\over 4y^2}+{1{-}{{\lambda}^2_\phi}\over 4(1+y)^2}+{2{+}K^2{-}{{\lambda}^2_\phi}{-}{{\lambda}^2_\psi}{-}a_f^2\tilde{\omega}^2\lambda\over4(1+y)}+{{-}2{-}K^2{+}{{\lambda}^2_\phi}{+}{{\lambda}^2_\psi}{+}a_f^2\tilde{\omega}^2(1{+}\lambda)\over 4y}\Bigg]\psi(y)=0
\ee
If we compare with the (2,0) quantum SW curve \cite{Bianchi:2021xpr,Bianchi:2021mft,Bonelli:2022ten,Consoli:2022eey}:
\be
\psi''(y)+\Bigg[{1-\left({m_1-m_2\over \hbar}\right)^2\over 4y^2}+{1-\left({m_1+m_2\over\hbar}\right)^2\over4(1+y)^2}+{1-2\left({m_1^2+m_2^2\over \hbar^2}\right)+{4u\over\hbar^2}\over 4(1+y)}+{-1+2\left(m_1^2+m_2^2\over \hbar^2\right)+{4(q-u)\over\hbar^2}\over 4y}\Bigg]\psi(y)=0
\ee
The dictionary between radial equation and $(2,0)$ q-SW curve reads:
\be
{m_{1,2}\over\hbar}={{\lambda}_\phi \mp \lambda_\psi \over 2} = {\mathcal{L}_\phi\mp \mathcal{L}_\psi\over 2a_f}\quad,\quad {u\over \hbar^2}={1-\tilde{\omega}^2(L_1^2+L_5^2)+K^2\over 4}\quad,\quad q={\hbar^2\tilde{\omega}^2 a_f^2\over4}
\ee
The angular equation in the coordinate $y=-\cos^2\q=-\xi$ and in canonical form becomes
\be
\psi''(y)+\Bigg[{2+K^2-m_\phi^2-m_\psi^2\over 4(1+y)}+{1-m_\psi^2\over 4y^2}+{1-m_\phi^2\over 4(1+y)^2}+{-2-K^2+a_f^2\tilde{\omega}^2+m_\phi^2+m_\psi^2\over 4y}\Bigg]\psi(y)=0
\ee
The dictionary between angular equation and $(2,0)$ q-SW reads
\be
{m_{1,2}^\q\over\hbar}={m_\phi\pm m_\psi\over2}\quad,\quad {u_\q\over\hbar^2}={1+K^2\over 4}\quad,\quad q_\q={\hbar^2\tilde{\omega}^2a_f^2\over 4}
\ee
Quite remarkably $q_\q=q_R$. 

Let us focus on the latter first. 
\subsubsection{Angular equation}
In the (2,0) flavour case the SW curve is given by
\be
P_L(x)=(x-m_{1,A})(x-m_{2,A})\quad,\quad P_R(x)=1\quad,\quad P_0(x)=x^2-u_A+q
\ee

The quantum period $a_A(u_A)$ can then be written as:
\begin{align}
a_A(u_A)&=\oint_\a \l_+=2\pi i\sum_{n=0}^\infty Res_{\sqrt{u_A}+n}\lambda_{+}(x)\\\nn
a_A(u_A)&=\sqrt{u_A}+{1-4m_{1,A}m_{2,A}-4u_A\over 4\sqrt{u_A}(4u_A-1)}q+\Bigg\{{2-3u_A\over64(u_A-1)u_A^{3/2}}+{3(m_{1,A}^2+m_{2,A}^2)\over16\sqrt{u_A}(1-5u_A+4u_A^2)}+\\\nn&+{m_{1,A} m_{2,A}(1-12u_A)\over4(1-4u_A)^2u_A^{3/2}}-{m_{1,A}^2m_{2,A}^2\Big[2+5u_A(12u_A-7)\Big]\over4(u_A-1)u^{3/2}(4u_A-1)^3}\Bigg\}q^2+\\\nn
&+\Bigg\{{5u_A-2-5u_A^2\over256(u_A-1)^2u_A^{5/2}}+{3(m_{1,A}^2+m_{2,A}^2)(1-15u_A+20u_A^2)\over64u_A^{3/2}(1-5u_A+4u_A^2)^2}+\\\nn&+{m_{1,A}m_{2,A}(m_{1,A}^2+m_{2,A}^2)(560u_A^3-1120u_A^2+497u_A-27)\over16(u_A-1)^2u_A^{3/2}(4u_A-9)(4u_A-1)^3}+\\\nn
&+{m_{1,A}m_{2,A}(54-861u_A+6397u_A^2-13100u_A^3+10480u_A^4-2880u_A^5)\over64(u_A-1)^2u_A^{5/2}(4u_A-9)(4u_A-1)^3}+\\\nn
&-{3m_{1,A}^2m_{2,A}^2(2-39u_A+371u_A^2-840u_A^3+560u_A^4)\over16(4u_A-1)^4(u_A-1)^2u_A^{5/2}}+\\\nn
&+{m_{1,A}^3m_{2,A}^3(18-413u_A+4705u_A^2-15260u_A^3+18480u_A^4-6720u_A^5)\over4(u_A-1)^2u_A^{5/2}(4u_A-9)(4u_A-1)^5}\Bigg\}q^3+{\cal O}(q^4)
\end{align}
Inverting this relation, one finds:
\begin{align}
\label{instseries}
&u_A=a_A^2+{4a_A^2+4m_{1,A}m_{2,A}-1\over2(4a_A^2-1)}q+\Bigg[{1\over 32(a_A^2-1)}-{3(m_{1,A}^2+m_{2,A}^2)\over 8-40 a_A^2+32a_A^4}+{(7+20a_A^2)m_{1,A}^2m_{2,A}^2\over2(a_A^2-1)(4a_A^2-1)^3}\Bigg]q^2+\\\nn
&+\Bigg[{5m_{1,A}m_{2,A}\over4(16a_A^6{-}56a_A^4{+}49a_A^2{-}9)}{-}{m_{1,A}m_{2,A}(m_{1,A}^2{+}m_{2,A}^2)(17{+}28a_A^2)\over(4a_A^2{-}1)^3(4a_A^4{-}13a_A^2{+}9)}{+}{4m_{1,A}^3m_{2,A}^3(29{+}232a_A^2{+}144a_A^4)\over(4a_A^2{-}1)^5(4a_A^4{-}13a_A^2{+}9)}\Bigg]q^3 \\
&\quad +{\cal O}(q^4)
\end{align}
The quantization condition for $a_A$ reads $a_A=n_A+{1\over2}$. Using the dictionary with the angular equation, we find:
\be
K^2= 4 u_A -1 = \ell(\ell+2)+4\sum_{k=1}^\infty u_k q^k
\ee
where we identified $n_A=\ell/2$ and $u_k$ are the coefficients of the instanton expansion for $u_A$ \eqref{instseries}.

\subsubsection{Radial equation}

The radial equation can be put in the standard form of a RCHE \cite{Bonelli:2022ten} 
$$
{d^2W\over dz^2}+ \left({\gamma\over z} + {\delta\over z-1}\right){dW\over dz} + {\beta z - \zeta \over z(z-1)} W = 0
$$
with gauge / Heun dictionary 
$$
u = \zeta - \beta + {1\over 4} (\gamma+\delta-1) \: , \quad q = \beta \:, \quad {\lambda}_\psi =\gamma    \:, \quad {\lambda}_\phi = \delta 
$$
Recall that $q_R=q_A= {1\over 4} \tilde\omega^2a_f^2$ while $u_R= u_A- {1\over 2} \tilde\omega^2L^2$ and ${\lambda}_\psi = m_\psi - P_z {L^2\over a_f}$, ${\lambda}_\phi = m_\phi - \omega {L^2\over a_f}$.

The connection from $z=0$ (regular) to $z=\infty$ (irregular) (e.g. $z=-\rho^2/a_f^2$) reads \cite{Bonelli:2022ten}
$$
\sqrt{\pi} z^{+{1\over 4} + {\gamma+\delta\over 2}} e^{\pm i \delta {\pi \over 2}} W^{RC}_0(\zeta,\beta,\gamma, \delta; z) = $$
$$  
\sum_{\sigma=\pm} A e^{i\pi \sigma(a-a_D)} e^{2i\sqrt{\beta z}} W^{RC}_\infty(\zeta,\beta,\gamma, \delta; {1\over\sqrt{z}}) + i \sum_{\sigma=\pm} A e^{-i\pi \sigma(a+a_D)} e^{-2i\sqrt{\beta z}} W^{RC}_\infty(\zeta,e^{2\pi i} \beta,\gamma, \delta; {1\over\sqrt{z}})
$$
where we used
$$
{a(\zeta)\Gamma(2\sigma a(\zeta))^2 
\Gamma(\gamma) (e^{i\pi} \beta)^{-{1\over 4} -\sigma a(\zeta)} e^{- {1\over 2} \partial_{a_0}{\cal F}  + {\sigma\over 2} \partial_{a}{\cal F}} \over \Gamma\left({\gamma-\delta +1 \over 2} +\sigma a(\zeta) \right) \Gamma\left({\gamma+\delta-1 \over 2} +\sigma a(\zeta) \right)} = A e^{i\pi \sigma(a-a_D)}
$$
with $A$ independent of $\sigma$, $2\pi i a_D = - \partial_a {\cal F}$ with ${\cal F}(a,q, m_f; \hbar)$ NS pre-potential and 
$$
W^{RC}_0(\zeta,\beta,\gamma, \delta; z) = 1 - {\zeta\over \gamma} z + {\gamma(\beta-\zeta) + \zeta(\zeta-\delta) \over 2\gamma(\gamma+1)} z^2 + {\cal O}(z^3)
$$
is the Reduced Confluent Heun function. 

All the relevant observables can be extracted from this formula. Prompt-ring down modes correspond to absence of the (ingoing) term $e^{-2i\sqrt{\beta z}}$ which is equivalent to the quantisation of the cycle $a_D-a$. In practice one has to compute the quantum period $a(u)$, invert it and use quantum Matone relation \cite{Matone:1995rx, Flume:2004rp} to find
${\cal F}(a,q, m_f; \hbar)$ and then $a_D= {-1\over 2\pi i} \partial {\cal F}/\partial a$. Expressing $a_D$ and $a$ in terms  of $u$, that in turn is given in terms of $u_A$ and $\tilde\omega L$, one finds an eigenvalue equation for $\omega_{QNM}$. The numerical results are collected in appendix \ref{appQNM}.

Due the absence of a horizon the absorption probability is zero. This is tantamount to saying that for real $\omega$ the reflected and incident waves from infinity have the same amplitude. Indeed
\begin{align}
&A_{I}=  i A \sum_{\sigma=\pm} \sigma e^{-i\pi \sigma(a+a_D)}  = 2 A \sin\pi(a+a_D)\\\nn
&A_{R}=  A \sum_{\sigma=\pm} \sigma e^{i\pi \sigma(a-a_D)}  = -2i A \sin\pi(a-a_D)
\end{align}
and for real $\omega$, $q=\omega^2a_f^2/4$ is real and so is $a$, while $a_D=i\tau$ is purely imaginary, thus
\begin{align}
{|A_{I}|^2\over 4|A|^2} =& |i\sinh\pi\tau\cos\pi a +\cosh\pi\tau\sin\pi a|^2 =  \\\nn
&=|-i\sinh\pi\tau\cos\pi a +\cosh\pi\tau\sin\pi a|^2= {|A_{R}|^2\over 4|A|^2} 
\end{align}

\subsubsection{Breit-Wigner resonance and direct integration Methods}
\label{BWresmetho}

The existence of a metastable minimum inside the photon-sphere implies the presence of a new class of quasi-bound states whose imaginary part, which is related to the tunnelling probability through the barrier, is in general very small. Since the imaginary part of the QNMs is related to the inverse of the life-time, we expect that this family of modes will dominate the late time of the ring-down phase (long-lived modes). As pioneered by Chandrasekhar and Ferrari \cite{ChanFerra} 
in the study of gravitational wave scattering on ultra-compact stars, quasi bound states will present as Breit-Wigner resonances in the scattering amplitude. More recently the resonance method has been used to compute long lived modes of AdS BHs \cite{Berti:2009wx} and of ultra-compact configurations like constant density stars and gravastars \cite{Cardoso:2014sna} whose effective potentials share astonishing analogies with the one for circular fuzz-ball, including the absence of a horizon. To implement these technique, our starting point is the choice of physically motivated boundary conditions.
The general asymptotic behaviour at infinity reads:
\be
R(\rho)\sim B(\omega) e^{-i \omega \rho}+C(\omega) e^{i \omega \rho}\qquad\qquad \rho\rightarrow \infty
\ee
The boundary condition $B=0$ describes purely outgoing wave at infinity {\it i.e.} prompt ring down modes, while the condition $C=0$ is associated to states localized in proximity of the minimum of the effective potential and exponentially decaying at spatial infinity, {\it i.e.} quasi-bound or long-lived modes.
Close to the `cap' at $\rho=0$, we impose regularity of the radial wave function:
\be
R(\rho)\sim \rho^{|m_\psi|}\qquad\qquad \rho\rightarrow 0
\ee
Starting from $\rho=0$, we numerically integrated until ``infinity'' $\rho=\Lambda >>L,a_f$ searching for exponentially decaying solutions ($C=0$). Then we proceeded by studying the minima of the square modulus of the numerically integrated radial function  which at this level ($\rho=\Lambda$) is a function $S(\omega)$ only of the frequency $\omega$, that may be considered (almost) real, since we expect $\omega_R>>|\omega_I|$. The minima of $|S(\omega)|^2$ will represent the real parts of long-live modes. Now, since the characteristic frequencies $\omega_0=\omega_R+i \omega_I$ are obtained by imposing $S(\omega)=0$ and since $\omega_R>>|\omega_I|$, we get
\be
S(\omega_R+i \omega_I)\simeq S(\omega_R)+i \omega_I {d S(\omega)\over d\omega}(\omega_R)=0
\ee
Expanding around $\omega_R$, restricting the argument of $S(\omega)$ to be real and using the previous expansion, we obtain:
\be
S(\omega)\simeq S(\omega_R)\left[1-{\omega-\omega_R\over i \omega_I}\sim \omega-\omega_R-i \omega_I \right]
\ee
So near the real $\omega$-axis:
\be\label{omegaI}
|S(\omega)|^2\simeq {\cal K} [(\omega-\omega_R)^2+\omega_I^2]
\ee
Using a parabolic fit near the minimum of $|S(\omega)|^2$ we extracted the value of the concavity ${\cal K}$, so we obtained a reliable approximation of $\omega_I$ by simply inverting \eqref{omegaI}. The whole procedure can be satisfactorily described as in Fig. \ref{BreitWignerfuzz}, while the numerical values of long-lived modes are collected in appendix \ref{appQNM}.
Direct integration method consists in numerically finding the complex roots of the function $S(\omega_0)=0$. Usually direct integration approach is very efficient and reliable in computing quasi-bound states since the boundary condition $C=0$ can be imposed on the leading behaviour of the radial function at infinity. Conversely prompt ring down modes, that would require $B=0$ as boundary condition, require to determine the subleading exponentially suppressed behaviour of radial function at infinity which could be contaminated by numerical error \cite{Pani:2013pma}.
\begin{figure}[h]
\centering
\includegraphics[width=8.5cm]{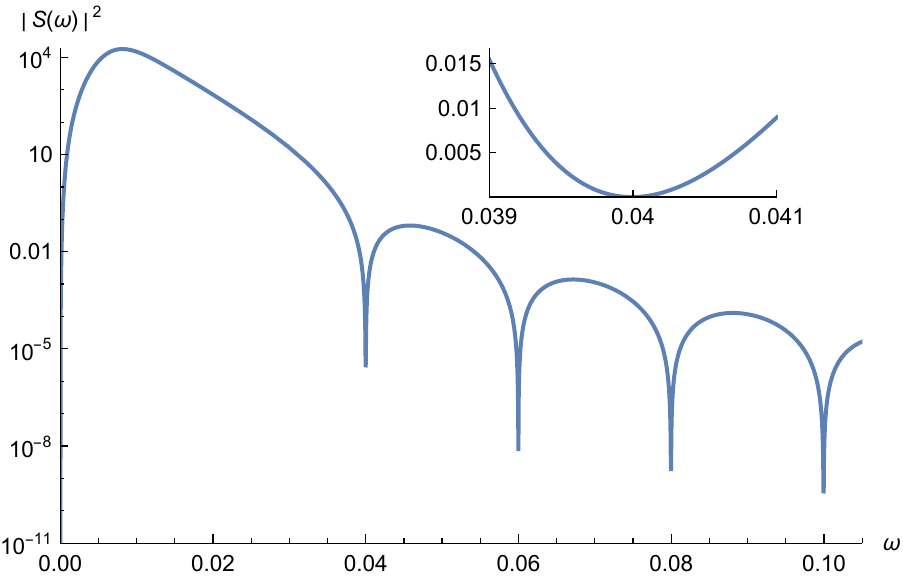}
\caption{Plot of the function $|S(\omega)|^2$ considered as a real function of $\omega$ for $a_f=0.01$ and angular quantum numbers $\ell=1$, $m_\phi=0$, $m_\psi=\pm1$.  Resonances are visible when $\omega_R\simeq0.04+n*0.02$. In the inset we plot the behaviour near the first minimum which shows a parabolic-like shape.}\label{BreitWignerfuzz}
\end{figure}

\subsection{Tidal Love numbers}

As for (small) BHs we can study the tidal Love number of circular fuzz-balls.

For static waves $\omega=0$ the problem becomes spherically symmetric, since the angular momentum of the fuzz-ball $a_f$  appears only in combination with $\omega$. As usual the tidal Love number is determined by the ratio of the coefficient of term decaying at infinity (`response') wrt the one growing at infinity (`source'). 

For $\omega=0$ and $\mu=0$ the radial equation in the variable $y=\rho^2/a_f^2$ is formally identical to the angular equation in the variable $z=\cos^2\theta$ up to a sign change $y=-z$ 
\be
\Psi''(z) + \left[ {1-m_\phi^2 \over 4(1-z)^2} + {1-m_\psi^2 \over 4z^2} - {K^2+2-m_\phi^2-m_\psi^2 \over 4z(z-1)} \right]\Psi(z) = 0
\ee
Setting $\Psi(z) = z^{1+|m_\psi| \over 2} (1-z)^{1+|m_\phi| \over 2} F(z)$ leads to a hypergeomtric equation with
\be
c= 1+|m_\psi| \quad , \quad a = {|m_\psi|+|m_\phi| - \ell\over 2} \quad 
b = 1+ {|m_\psi|+|m_\phi| + \ell\over 2} = a+\ell +1
\ee 
Notice that
\be
\ell = |m_\psi|+|m_\phi| + 2r
\ee
with $r=0,1,2,....$ ($r=0$ is the wave counterpart of the shear-free geodesics) so that $a=-r$ and $b= \ell - r+ 1\ge r+1$ and $c$ is an integer.   
 
The `regular' solution (at $z=0$ and $z=1$) of the angular equation are the spherical harmonics on $S^3$
\be 
S_\ell(\theta)e^{i m_\phi \phi} e^{i m_\psi \psi}= D^{(\ell/2)}_{m_1, m_2} (2\theta, \phi, \psi) = d^{(\ell/2)}_{m_1, m_2} (2\theta) e^{i (m_1+m_2)\phi} e^{i (m_1-m_2)\psi}
\ee
which in turn are related to the characters of $SU(2)$ that are polynomial in $z$. 

Using the relation with the hypergeometric equation, the `regular' solution reads
\be
\Psi_{reg}(z) = z^{1+|m_\psi| \over 2} (1-z)^{1+|m_\phi| \over 2} F(-r, \ell-r+1; 1+|m_\psi|, z)
\ee
with $F_r(z)=F(-r, \ell-r+1; 1+|m_\psi|, z)$ a degree $r$ (Jacobi) polynomial in $z=-\rho^2/a_f^2$. The other independent solution is irregular at both $z=0$ (`cap') and $z=1$ (unphysical) since it develops a logarithmic behaviour\footnote{Indeed the second solution is of the form $G(z)= \log(z) F_r(z)+ \sum_{k=-m_\psi}^{\infty} c_k z^{k}$.} due to the integer difference between the Frobenius exponents: $\Delta \alpha_0 = |m_\psi|$ and $\Delta \alpha_1 = |m_\phi|$.

Going back to the `physical' perturbation described by $R(\rho) = {\Psi\over\sqrt{z(1-z)}}$ 
and imposing regularity at $z=0$, {\it i.e.} choosing $R_{reg}(z) = {\Psi_{reg}(z)\over\sqrt{z(1-z)}}$ 
fixes completely the behaviour at $\infty$
 \be
 R_{reg}(z) \sim z^{r + {|m_\phi| + |m_\phi| \over 2}} \left(1-{1\over z}\right)^{|m_\phi| \over 2} {F(-r, \ell-r+1; 1+|m_\psi|, z) \over z^r} = z^{{\ell\over 2}} + ... + \kappa^\ell_{m_\phi, m_\psi} z^{-{\ell\over 2}-1}
 \ee
 where  $\kappa^\ell_{m_\phi, m_\psi}$ is the coefficient of the term of order $z^{-\ell-1}$ in the expansion of the factor 
 \be f(z)=(1-{1\over z})^{|m_\phi| \over 2} {F(-r, \ell-r+1; 1+|m_\psi|, z) \over z^r}
 \ee
One easily finds
\be
\kappa^\ell_{m_\phi, m_\psi} = 
{(-1)^{m_{\psi}+m_{\phi}+r+1} \Gamma \left(1+\frac{|m_{\phi}|}{2}\right) \over
\Gamma\left(-r- |m_{\psi}| - \frac{|m_{\phi}|}{2}\right) \Gamma (|m_{\phi}|+|m_{\psi}|+r+2}\times
\ee
$${}_3F_2\left(-r,\frac{|m_{\phi}|}{2}+|m_{\psi}|+r+1,|m_{\phi}|+|m_{\psi}|+r+1; |m_{\psi}|+1,|m_{\phi}|+|m_{\psi}|+r+2;1\right)
$$
that looks finite and non-zero at least for $m_\phi$ odd, quite independently of the value of $m_\psi$, while it vanishes for $m_\phi$ even.  

More remarkably, $\kappa^\ell_{m_\phi, m_\psi}$ is totally independent of $a_f$ and  $L$ so even for `typical' fuzz-balls with $a_f<<L$ one would have a non-zero result (for $m_\phi $ odd)\footnote{We thank Daniel Mayerson for enlightening discussion on this point.}.

It is often customary to define TLN from an asymptotic expansion in terms of the rescaled radial variables $\hat{r} = r/(r_+{-}r_-)$ or $\tilde{r} = r/2M$ in $D=4$ or $\hat{r}^2 = r^2/(r_+^2{-}r_-^2)$ or $\tilde{r}^2 = r^2/2M$ in $D=5$. In our case this would suggest introducing $\tilde\rho= \rho L/a$. As a result the TLN would rescale as $\tilde\kappa^\ell_{m_\phi, m_\psi} = \kappa^\ell_{m_\phi, m_\psi} a_f^{2\ell+2} / L^{2\ell+2}$ and would smoothly go to zero as $a_f\rightarrow 0$.

Anyway it seems that contrary to small BHs
\be
{\cal L}^{fuzz}_{stat} \neq 0 
\ee 
Also notice that contrary to `standard' $D=5$ BHs \cite{Pereniguez:2021xcj}\footnote{Similar results have been obtained by Daniel Mayerson for BMPV BHs. We thank him for sharing his insights.}  no log terms are possible since they would be incompatible with regularity at $z=0$ and they cannot be generated by analytic continuation of $R_{reg}(z) = z^{|m_\psi| \over 2} (1-z)^{|m_\phi| \over 2} F_r$
to infinity since $F_r$ is a polynomial. This is another feature that distinguishes (circular) fuzz-balls from (small) BHs even in the rigorously static limit.

\begin{table}[]
\centering
\begin{tabular}{|c|c|c|}
\hline
                       & Angular quantum numbers       & Static Love Number \\ \hline
$\ell=0$               & $m_\phi=0$, $m_\psi=0$, $r=0$ & 0                  \\ \hline
$\ell=1$               & $m_\phi=1$, $m_\psi=0$, $r=0$ & $-{1\over8}$       \\ \cline{2-3}
                       & $m_\phi=1$, $m_\psi=1$, $r=0$ & 0                  \\ \hline
$\ell=2$               & $m_\phi=2$, $m_\psi=0$, $r=0$ & 0                  \\ \cline{2-3}
                       & $m_\phi=1$, $m_\psi=1$, $r=0$ & $-{1\over16}$      \\ \cline{2-3}
                       & $m_\phi=0$, $m_\psi=2$, $r=0$ & 0                  \\ \cline{2-3}
                       & $m_\phi=0$, $m_\psi=0$, $r=1$ & 0                  \\ \hline
$l=3$                  & $m_\phi=3$, $m_\psi=0$, $r=0$ & ${3\over128}$      \\ \cline{2-3}
                       & $m_\phi=2$, $m_\psi=1$, $r=0$ & 0                  \\ \cline{2-3}
                       & $m_\phi=1$, $m_\psi=2$, $r=0$ & $-{5\over128}$     \\ \cline{2-3}
                       & $m_\phi=0$, $m_\psi=3$, $r=0$ & 0                  \\ \cline{2-3}
\multicolumn{1}{|l|}{} & $m_\phi=1$, $m_\psi=0$, $r=1$ & ${7\over128}$      \\ \cline{2-3}
\multicolumn{1}{|l|}{} & $m_\phi=0$, $m_\psi=1$, $r=1$ & 0                  \\ \hline
\end{tabular}
\caption{Some TLNs for different choices of angular quantum numbers. Note that TLNs are non-zero for $m_\phi$ odd.}\label{TLNfuzz}
\end{table}

This non-zero result in the `rigorously' static case should be compared with the $\omega\rightarrow 0$ limit of the `dynamical' Love number with $\omega \neq 0$ via the `Love function' (defined for $L<<r<<1/\omega$) by imposing regularity at the horizon (or at the `cap') and then taking the ratio of the coefficient of the `response' term decaying at infinity (as $\rho^{-\ell - 2 - \Delta\ell (\omega)}$) and of the `source' term growing at infinity (as $\rho^{+\ell + \Delta\ell(\omega)}$). Once again and for the same reasons as before, the two limits in general do not commute  \cite{Consoli:2022eey}. We collect some values of the TLNs for different choises of angular quantum numbers in Table \ref{TLNfuzz}

\section{Conclusions and outlook}

We have studied scalar perturbations of circular fuzz-balls and compared the results for physical observable such as QNMs, TLNs and grey-body factor with the ones for a small BH obtained in the limit $a_f\rightarrow 0$. 

Let's summarise our results:
\begin{itemize}
\item The circular fuzz-ball presents two branches of QNMs: the prompt-ring down, a.k.a. photon-sphere, modes localised at the maximum of the radial effective potential and long-lived modes, a.k.a. quasi-bound modes, localised at a local (metastable) minimum of the radial effective potential; small BH only admit prompt-ring down modes that match with those of the circular fuzz-ball when $a_f<<L$. 
\item The circular fuzz-ball has zero absorption cross section, due to the absence of a horizon. On the contrary, small BHs, very much as D3-branes \cite{Gubser:1998iu} have a non-zero $\sigma_{abs}(\omega)$ notwithstanding the vanishing area of the horizon. We computed $P_{abs}(\omega)$ and then $\sigma_{abs}(\omega)$ relying on a combination of properties of (modified) Mathieu equation, that is the qSW curve for pure ${\cal N}=2$ SYM theory, and instanton techniques \`a la Nekrasov-Shatashvili.
\item Even in the rigorously static case $\omega =0$ the circular fuzz-ball expose non-zero TLNs when one of the two azimuthal `quantum' numbers ($m_\phi$) is odd. The (adimensional) results are independent of $a_f$ and $L$. A rescaling by $(a_f/L)^{2\ell+2}$ is suggested by similar analyses for BHs.  On the contrary small BHs have zero TLNs in the rigorously static case, while the static limit $\omega\rightarrow 0$ of the Tidal Love function, introduced in \cite{Consoli:2022eey} produces a non-zero results. 
\item In both cases absence of an ergo-region (since $g_{tt}>0$ always and everywhere) prevents super-radiance to take place. Yet charge super-radiance is expected to take place at the small BH `horizon' for probes with charge $q$ when $\omega < q \Phi_{H}$, with $\Phi_{H}$ the `electric' potential at the horizon\footnote{We plan to study this and similar effects in the near future \cite{MBCdBGdR}.} 
\end{itemize}

As expected and in some sense required for a correct description of the small BH `micro states', the differences tend to disappear for small $a_f$, except for the adimensional TLNs. Yet the absence of any absorption that is replaced by the presence of echoes, associated to long-lived quasi-bound modes, produces observable features that should  allow to `classically' discriminate fuzz balls from BHs. In particular `ensemble average' may lead to measurable deviations in the GW signal of fuzz-ball mergers as compared with BH mergers. These can be hinted at by studying the evolution in the time domain of a gaussian shell along the lines of \cite{Ikeda:2021uvc} in the single micro-state or `averaged' fuzz-ball.  

For simplicity we worked with (massive) scalar perturbations, while the ultimate goal would be to study GW or EM waves. However in the BPS context we performed our analysis, residual supersymmetry transformation should allow to map spin 0 to spin 1 and possibly spin 2 and reduce the problem to the correct identification of the Killing spinors. This applies to more realistic micro-state geometries based on multi-center ansatze that admit only reduced isometry, whereby vector or tensor perturbations would really be hard to analyse. 

Our approach is largely based on integrability so it is hard to apply to such micro states. One aspect that could still be addressed is the breaking of equatorial symmetry emphasised in \cite{Fransen:2022jtw} that is present in a class of BHs, known as Rasheed-Larsen BHs \cite{Larsen:1999pp, Rasheed:1995zv}, that lead to separable dynamics. It would be interesting to apply our techniques in this or similar cases. 

Integrability is known to take place also in some 3-charge superstrata \cite{Bena:2016ypk} with AdS asymptotics, corresponding to a decoupling limit that only focus on the `cap' region. A preliminary analysis shows that this description could only capture the long-lived / quasi-bound modes, that are in fact exactly stable in this approximation due to the infinite wall at the boundary of AdS, but not the  prompt ring-down modes. Establishing a connection between TLNs as response to a source and response functions in the AdS context \cite{Bena:2019azk} could elucidate the meaning of the non-vanishing adimensional TLNs that we found here. Last but not least  one would like to explore similar properties of neutral micro-states (possibly with `large' dipole moments) \cite{Bah:2022yji, Mayerson:2022ekj, Ganchev:2022vrv}.

\section*{Acknowledgements}

We would like to thank P.~Arnaudo, G.~Bonelli, G.~Bossard, D.~Capocci, D.~Consoli,  M.~Firrotta, F.~Fucito,  A.~Grassi, A.~Grillo, A.~Hashimoto, C.~Iossa, D.~Mayerson, J.~F.~Morales, N.~Nekrasov, D.~Panea-Lichtig, P.~Pani, R.~Poghosyan,  F.~Riccioni, R.~Savelli, A.~Tanzini,  for interesting discussions and clarifications. 
M.~B. would like to thank Ecole Polytechnique for the kind hospitality while this work was being completed. We thank the MIUR PRIN contract $2020KR4KN2$ String Theory as a bridge between Gauge Theories and Quantum Gravity and the INFN project $ST\&FI$ String Theory and Fundamental Interactions for partial
support.

\appendix

\section{The AGT picture}
\label{AGT}

AGT duality \cite{Alday:2009aq} relates 4-d $\mathcal{N} = 2$ quiver theories  in a non-commutative $\Omega$-background \`a la Nekrasov-Shatasvili \cite{Nekrasov:2009rc, MM0910.5670, Z1103.4843} to 2-d  Liouville CFT. 
Denoting by ${\epsilon_1},{\epsilon_2}$ the quantum deformation parameters and by 
$b = \sqrt{\frac{\epsilon_1}{\epsilon_2}}$, the central charge of the CFT is given by 
\begin{equation*}
c = 1 + 6 Q^2
\quad {\rm with} \quad
Q = b + \frac{1}{b}
\end{equation*}
AGT relates conformal blocks of the 2-d  Liouville CFT to (ratioes of) quiver partition functions
\begin{equation*}
\mathcal{C}_{p_0\cdots p_{n+1}}^{\alpha_1 \cdots \alpha_{n+1}}\left(\{z_i\}\right)\prod_{j=1}^n z_j^{-\Delta_{p_j} + \Delta_j + \Delta_{p_j+1}} = \frac{Z_\text{inst}\left(\{\vec{a}_i\},\{q_i\}\right)}{Z_{U(1)}(\{q_i\})}
\end{equation*}
the dictionary relates insertion points $z_i$ to (ratios of) gauge couplings $q_i$, the dimension $\Delta_i$ of the chiral vertex operators to masses $m_f$ or scalar vev's $a_r$. Finally the BPZ (Belavin-Polyakov-Zamolodchikov) equation satisfied by the conformal blocks is related to the quantum SW curve, which in turn plays the role of wave equation in a BH or fuzz-ball background, the central theme of our analysis. 

In order to briefly illustrate the procedure\footnote{See \cite{Consoli:2022eey} for a more detailed description in similar contexts.}, consider the $n+3$-point function of chiral vertex operators $V_{\alpha_i} = e^{2\alpha_i \phi}$ with $\Delta_i = \alpha_i\left(Q - \alpha_i\right)$
\begin{equation*}
{\cal G}(z_i,\bar{z}_i) = \langle \prod_{i = 0}^{n+2} V_{\alpha_i}(z_i)\rangle 
\end{equation*}
and expand it in terms of conformal blocks
\begin{equation*}
{\cal G}(z_i,\bar{z}_i) = \sum_{p_1\cdots p_n} \langle p_0|V_{\alpha_0}(z_0)|p_1\rangle \cdots \langle p_n|V_{\alpha_{n+2}}(z_{n+2})|p_{n+1}\rangle 
= \sum_{p_1\cdots p_n}\left|\mathcal{C}_{p_0\cdots p_{n+1}}^{\alpha_1 \cdots \alpha_{n+1}}\right|^2
\end{equation*}
For definiteness choose $n=2$, related to an $SU(2)\times SU(2)$ quiver. The NS partition function, to be identified with the wave-function, is associated to conformal block involving level 2 degenerate field with $\alpha_3 = - {b}/{2}$
$$
\Psi(\{z_i\})= \mathcal{C}_{p_0\cdots p_{3}}^{\alpha_1 \cdots \alpha_{3}}\left(\{z_i\}\right)
$$
Since $\left(L_{-1}^2 + b^2 L_2\right)V_{\alpha_3} \sim 0$ is null, $\Psi(\{z_i\})$ satisfies  the BPZ equation
$$
\Psi''(\{z_i\}) + b^2 \sum_{i \neq 1}^{4}\left[\frac{\Delta_i}{(z-z_i)^2} + \frac{1}{z - z_i} \partial_{z_i}\right] \Psi(\{z_i\}) = 0
$$
Setting $z_0 = \infty$, $z_1 = 1$, $z_2 = q$, $z_3 = z$, $z_4 = 0$ and performing a semi-classical double scaling limit $b\rightarrow 0$, {\it i.e.} $\epsilon_1=0$ in the NS $\Omega$-background, with fixed  
$$
b^2 \Delta_i = \delta_i
\,,\quad
b^2 c_i = \nu_i 
\,,\quad
\partial_{z_i} \Psi(y;\{z_i\}) = c_i \Psi(y;\{z_i\})
$$
one gets the quantum SW curve / differential equation for $\Psi(z)$ with\footnote{There are many possible choices of signs in the following expressions.} 
\begin{align}\label{dict1}
\d_{p_0}={1\over 4}-p_0^2\quad,&\quad p_0={m_1-m_2\over 2}\\\nn
\d_{k_0}={1\over4}-k_0^2\quad,&\quad k_0={1+m_1+m_2\over2}\\\nn
\d_{p_3}={1\over 4}-p_3^2\quad,&\quad p_3={m_3-m_4\over 2}\\\nn
\d_{k}={1\over4}-k^2\quad,&\quad k={1-m_3-m_4\over2}\\\nn
c=k+p_3 \quad,&\quad c' = p_0-k_0
\end{align}

\section{Tables}\label{appQNM}

\begin{table}[h!]
\centering
\begin{tabular}{|c|c|c|}
\hline
$n=0$      & WKB & Seiberg-Witten     \\ \hline
$a_f=0$ (BH)& 0.433013-0.25 I   &  0.498509 - 0.278493 I  \\\hline
$a_f=0.01$ & 0.433037 - 0.249966 I    &  0.498493 - 0.263181 I      \\ \hline
$a_f=0.02$ &  0.43311 - 0.249862 I   &  0.498476 - 0.263177 I      \\ \hline
$a_f=0.05$ & 0.433622 - 0.249139 I    &  0.498352 - 0.263152 I    \\ \hline
$a_f=0.1$  &  0.435459 - 0.246543 I   &  0.497917 - 0.263058 I      \\ \hline
$a_f=0.15$ &  0.438544 - 0.242183 I   &  0.497212 - 0.262893 I       \\ \hline
$a_f=0.2$  &   0.442909 - 0.236058 I  &   0.496261 - 0.262649 I      \\ \hline
$a_f=0.25$ &  0.448583 - 0.228314 I   &   0.495096 - 0.262321 I    \\ \hline
$a_f=0.3$  &  0.455564 - 0.219488 I   &  0.493746 - 0.261908 I     \\ \hline
\end{tabular}
\caption{Prompt ringdown modes : $l=0\Rightarrow m_\phi=m_\psi=0$}
\end{table}
\vspace{1cm}
\begin{table}[h!]
\centering
\begin{tabular}{|c|c|c|}
\hline
$n=0$      & WKB & Seiberg-Witten   \\ \hline
$a_f=0$ (BH) &  0.968246-0.25 I  &   0.999627 - 0.255004 I\\\hline
$a_f=0.01$ &  0.968272 - 0.249978 I   &  0.999742 - 0.253996 I    \\ \hline
$a_f=0.02$ &  0.968352 - 0.249912 I   &   0.999769 - 0.253954 I     \\ \hline
$a_f=0.05$ &  0.968913 - 0.249446 I   &   0.999962 - 0.253661 I    \\ \hline
$a_f=0.1$  &   0.970928 - 0.247753 I  &    1.00064 - 0.252607 I   \\ \hline
$a_f=0.15$ &  0.974335 - 0.244826 I   &   1.00176 - 0.250821 I    \\ \hline
$a_f=0.2$  &  0.979216 - 0.240485 I   &   1.00326 - 0.248265 I      \\ \hline
$a_f=0.25$ &  0.9857 - 0.234407 I   &    1.0051 - 0.244889 I    \\ \hline
$a_f=0.3$  &  0.993986 - 0.226022 I   &   1.00719 - 0.240641 I \\ \hline
\end{tabular}
\caption{Prompt ringdown modes: $l=1\Rightarrow m_\phi=0, m_\psi=\pm1$}
\end{table}
\vspace{1cm}
\begin{table}[h!]\small
\centering
\begin{tabular}{|c|c|c|c|}
\hline
$n=0$      & WKB                            & Direct Integration & Breit-Wigner            \\ \hline
$a_f=0.01$ & $0.0380185{-}7.57895E{-}15 I$ & $0.039992{-}(****)E{-}14 I$  & $0.039992{-}(****)I$ \\ \hline
$a_f=0.02$ & $0.0759967{-}3.34991E{-}12 I$ & $0.079936{-}2.06087E{-}12 I$  &$0.079936{-}3.31768E{-}12I$ \\ \hline
$a_f=0.05$ & $0.18929{-}1.17547E{-}8 I$    & $0.199001{-}8.11612E{-}10 I$  & $0.199001{-}8.39096E{-}10I$ \\ \hline
$a_f=0.1$  & $0.373622{-}7.24191E{-}6 I$   & $0.392034{-}5.38521E{-}7 I$  &  $0.392034{-}5.86285E{-}7I$  \\ \hline
$a_f=0.15$ & $0.548311{-}0.000376738 I$     & $0.573191{-}0.0000267682 I$  &  $0.573191{-}0.0000288871 I$   \\ \hline
$a_f=0.2$  & $0.709157{-}0.00680183 I$      & $0.73646{-}0.000489566 I$  & $0.73646{-}0.000490324 I$    \\ \hline
$a_f=0.25$ & $0.852547{-}0.0643601 I$       & $0.875631{-}0.00358531 I$  & $0.876019{-}0.00362734 I$    \\ \hline
\end{tabular}
\caption{Long-lived modes modes: $l=1\Rightarrow m_\phi=0, m_\psi=\pm1$}
\end{table}
\vspace{1cm}
\begin{table}[h!]
\centering
\begin{tabular}{|c|c|c|}
\hline
$n=0$      & WKB & Seiberg-Witten \\ \hline
$a_f=0$ (BH) &  0.968246-0.25 I  &   0.999627 - 0.255004 I\\\hline
$a_f=0.01$ &  0.970767 - 0.249338 I   &  1.00222 - 0.253381 I    \\ \hline
$a_f=0.02$ &  0.973331 - 0.248641 I   &   1.00471 - 0.252735 I     \\ \hline
$a_f=0.05$ &  0.981289 - 0.246332 I   &   1.01225 - 0.250681 I    \\ \hline
$a_f=0.1$  &  0.995494 - 0.241681 I   &  1.02498 - 0.246864 I    \\ \hline
$a_f=0.15$ &  1.01099 - 0.235877 I   &   1.03789 - 0.242534 I  \\ \hline
$a_f=0.2$  &  1.0279 - 0.22874 I   &  1.05093 - 0.237671 I   \\ \hline
$a_f=0.25$ &   1.04637 - 0.220116 I  &  1.06405 - 0.23227 I     \\ \hline
$a_f=0.3$  &  1.06652 - 0.209993 I   &   1.0772 - 0.226339 I     \\ \hline
\end{tabular}
\caption{Prompt ringdown modes: $l=1\Rightarrow m_\phi=1, m_\psi=0$}
\end{table}
\vspace{1cm}
\begin{table}[h!]
\centering
\begin{tabular}{|c|c|c|c|}
\hline
$n=0$      & WKB & Seiberg-Witten   \\ \hline
$a_f=0$ (BH) &  0.968246-0.25 I  &   0.999627 - 0.255004 I\\\hline
$a_f=0.01$ &  0.965767 - 0.250629 I   &   $0.997258 - 0.25462$ I   \\ \hline
$a_f=0.02$ &  0.963329 - 0.251226 I   &   $0.994793 - 0.255212$ I   I   \\ \hline
$a_f=0.05$ &  0.956253 - 0.252833 I   &   $0.98746 - 0.256884$ I     \\ \hline
$a_f=0.1$  &  0.94521 - 0.254956 I   &   $0.97545 - 0.259339$ I   \\ \hline
$a_f=0.15$ &  0.935027 - 0.256475 I   &   $0.963716 - 0.261413$ I   \\ \hline
$a_f=0.2$  &   0.925625 - 0.257482 I  &   $0.952263 - 0.263141$ I      \\ \hline
$a_f=0.25$ & 0.916931 - 0.258053 I    &   $0.941093 - 0.264559$ I      \\ \hline
$a_f=0.3$  &  0.908883 - 0.258258 I   &  $0.930206 - 0.265698$ I     \\ \hline
\end{tabular}
\caption{Prompt ringdown modes: $l=1\Rightarrow m_\phi=-1, m_\psi=0$}
\end{table}

\begin{table}[h!]
\centering
\begin{tabular}{|c|c|c|}
\hline
$n=0$      & WKB & Seiberg-Witten   \\ \hline
$a_f=0$ (BH) &  1.47902-0.25 I  &   1.50114 - 0.250172 I\\\hline
$a_f=0.01$ &  1.47904 - 0.249988 I   &   1.49856 - 0.24844 I     \\ \hline
$a_f=0.02$ &  1.47908 - 0.249954 I   &   1.49857 - 0.248423 I   \\ \hline
$a_f=0.05$ &  1.47941 - 0.249712 I   &   1.49862 - 0.248304 I    \\ \hline
$a_f=0.1$  &  1.48058 - 0.248846 I   &   1.49881 - 0.247884 I     \\ \hline
$a_f=0.15$ &  1.48252 - 0.247396 I   &  1.49913 - 0.247201 I     \\ \hline
$a_f=0.2$  &  1.48525 - 0.245354 I   &   1.4996 - 0.24628 I      \\ \hline
$a_f=0.25$ &  1.48874 - 0.242712 I   &  1.50026 - 0.245149 I   \\ \hline
$a_f=0.3$  &  1.49301 - 0.239465 I   &   1.50114 - 0.243834 I    \\ \hline
\end{tabular}
\caption{Prompt ringdown modes: $l=2\Rightarrow m_\phi=m_\psi=0$: the first non-shear free}
\end{table}

\begin{table}[h!]
\centering
\begin{tabular}{|c|c|c|}
\hline
$n=0$      & WKB & Seiberg-Witten    \\ \hline
$a_f=0$ (BH) &  1.47902-0.25 I  &   1.50114 - 0.250172 I\\\hline
$a_f=0.01$ &  1.47906 - 0.249975 I   &   $1.49859 - 0.248427$ I     \\ \hline
$a_f=0.02$ &  1.4792 - 0.249902 I   &   $1.49868 - 0.248369$ I     \\ \hline
$a_f=0.05$ &  1.48014 - 0.249383 I   &    $1.49933 - 0.247964$ I    \\ \hline
$a_f=0.1$  &  1.48355 - 0.247482 I   &   $1.50165 - 0.246489$ I    \\ \hline
$a_f=0.15$ &  1.48935 - 0.244131 I   &  $1.50552 - 0.243924$ I I  \\ \hline
$a_f=0.2$  &  1.49779 - 0.238996 I   &    $1.51094 - 0.240099$ I   \\ \hline
$a_f=0.25$ & 1.50922 - 0.231446 I   &  $1.5179 - 0.234761$ I      \\ \hline
$a_f=0.3$  &  1.52428 - 0.22021 I   &   $1.52629 - 0.227565$ I   I \\ \hline
\end{tabular}
\caption{Prompt ringdown modes: $l=2\Rightarrow m_\phi=0, m_\psi=\pm 2$}
\end{table}
\vspace{1cm}
\begin{table}[h!]\small
\centering
\begin{tabular}{|c|c|c|c|}
\hline
$n=0$      & WKB & Direct Integration & Breit-Wigner \\ \hline
$a_f=0.01$ & $0.0589341{-}3.59432E{-}21 I$   &  $0.0599879{-}(****)I$  &   $0.0599879-(****)I$   \\ \hline
$a_f=0.02$ &  $0.117801{-}2.74315E{-}17 I$  &  $0.119904{-}(****)I$  &    $0.119904-(****)I$  \\ \hline
$a_f=0.05$ & $0.293332{-}4.42517E{-}12 I$ &  $0.298503{-}(****)E{-}9 I$ &  $0.298503-1.80738E{-}9 I$   \\ \hline
$a_f=0.1$  & $ 0.578388{-}5.65627E{-}8 I$ &  $0.588085{-}4.99815E{-}8 I$  &  $0.588085-5.23269E{-}8I$   \\ \hline
$a_f=0.15$ & 0.847314{-}0.0000191774 I & $0.8601{-}8.91*10^{-7} I$   &   $0.8601-9.37417E{-}7 I$    \\ \hline
$a_f=0.2$  & 1.093{-}0.00135584 I   &  $1.10614{-}0.0000603124 I$ &   $1.10614-0.0000698527 I $     \\ \hline
$a_f=0.25$ &  1.30917{-}0.0361528 I & $1.31713{-}0.0013159 I$   &  $1.31716-00135213 I$   \\ \hline
\end{tabular}
\caption{Long-lived modes: $l=2\Rightarrow m_\phi=0, m_\psi=\pm 2$}
\end{table}
\vspace{1cm}
\begin{table}[h!]\small
\centering
\begin{tabular}{|c|c|c|}
\hline
$n=0$      & WKB & Seiberg-Witten   \\ \hline
$a_f=0$ (BH) &  1.47902-0.25 I  &   1.50114 - 0.250172 I\\\hline
$a_f=0.01$ &  1.48155 - 0.249561 I   &  $1.50118 - 0.247959$ I   \\ \hline
$a_f=0.02$ &  1.48413 - 0.249091 I   &  $1.50382 - 0.247455$ I  \\ \hline
$a_f=0.05$ & 1.49219 - 0.247476 I    &  $1.51193 - 0.245834$ I    \\ \hline
$a_f=0.1$  &  1.50678 - 0.244051 I   &  $1.52602 - 0.242725$ I        \\ \hline
$a_f=0.15$ &  1.52294 - 0.239567 I   &  $1.54077 - 0.239004$ I         \\ \hline
$a_f=0.2$  &  1.54084 - 0.233792 I  &  $1.55616 - 0.234501$ I      \\ \hline
$a_f=0.25$ & 1.56074 - 0.226362 I  &  $1.57214 - 0.228992$ I       \\ \hline
$a_f=0.3$  &  1.58298 - 0.216656 I  &  $1.58864 - 0.222206$ I     \\ \hline
\end{tabular}
\caption{Prompt ringdown modes: $l=2\Rightarrow m_\phi=1, m_\psi=\pm 1$}
\end{table}
\vspace{1cm}
\begin{table}[h!]\small
\centering
\begin{tabular}{|c|c|c|c|}
\hline
$n=0$      & WKB & Direct Integration  &  Breit-Wigner \\ \hline
$a_f=0.01$ & $0.05823{-}2.5985E{-}23$ I  &  $0.059988{-}(****)I$  &  $0.059988{-}(****) I$      \\ \hline
$a_f=0.02$ & $0.116394{-}1.97546E{-}19$ I     &  $0.119904{-}(****)I$  & $0.119904{-}(****) I$    \\ \hline
$a_f=0.05$ & $0.289845{-}3.12586E{-}14$ I  &  $0.298504{-}(****)I$  &  $0.298504{-}(****) I$     \\ \hline
$a_f=0.1$  & $0.571628{-}3.81488E{-}10$ I  &  $0.588116{-}1.101E{-}9 I$  &  $0.588116{-}1.59509E{-}9 I$       \\ \hline
$a_f=0.15$ & $0.837718{-}1.22639E{-7}$ I &  $0.860343{-}3.91098E{-}7 I$   &  $0.860343{-}4.06279E{-}7 I$     \\ \hline
$a_f=0.2$  &  $1.081246{-}8.20825E{-}6$I &  $1.10727{-}0.0000262723 I$  &  $1.10727{-}0.0000278005 I$      \\ \hline
$a_f=0.25$ & $1.296233{-}0.000206853$ I  & $1.32108{-}0.000633244 I$  &   $1.32108{-}0.000633721 I$      \\ \hline
\end{tabular}
\caption{Long-lived modes: $l=2\Rightarrow m_\phi=1, m_\psi=\pm 1$}
\end{table}
\vspace{1cm}
\begin{table}[h!]
\centering
\begin{tabular}{|c|c|c|c|}
\hline
$n=0$      & WKB & Seiberg-Witten \\ \hline
$a_f=0$ (BH) &  1.47902-0.25 I  &   1.50114 - 0.250172 I\\\hline
$a_f=0.01$ &  1.47655 - 0.250407 I   &   1.49598 - 0.248916 I    \\ \hline
$a_f=0.02$ & 1.47412 - 0.250783 I   &  1.49342 - 0.24937 I      \\ \hline
$a_f=0.05$ &  1.46716 - 0.251731 I  &  1.48594 - 0.250635 I     \\ \hline
$a_f=0.1$  &  1.45652 - 0.252742 I   &  1.47411 - 0.252442 I   \\ \hline
$a_f=0.15$ &  1.44703 - 0.25308 I  &   1.46308 - 0.253881 I    \\ \hline
$a_f=0.2$  & 1.43864 - 0.252765 I   &  1.45287 - 0.254947 I \\ \hline
$a_f=0.25$ & 1.43132 - 0.251796 I  &  1.4435 - 0.255624 I \\ \hline
$a_f=0.3$  & 1.42503 - 0.250146 I   & 1.43497 - 0.255878 I \\ \hline
\end{tabular}
\caption{Prompt ringdown modes: $l=2\Rightarrow m_\phi=-1, m_\psi=\pm 1$}
\end{table}
\clearpage
\begin{table}[h!]\small
\centering
\begin{tabular}{|c|c|c|c|}
\hline
$n=0$      & WKB & Direct integration &  Breit-Wigner  \\ \hline
$a_f=0.01$ &  $0.0382363{-}3.25139E{-}24$ I   & $0.0399947{-}(****)$ I  &  $0.0399947{-}(****) I$      \\ \hline
$a_f=0.02$ & $0.0764463{-}2.41336E{-}20$ I    &  $0.0799573{-}(****)$ I&  $0.0799573{-}(****) I$  \\ \hline
$a_f=0.05$ & $0.190657{-}3.37968E{-}15$ I  & $0.199335{-}(****)$ I   &   $0.199335{-}(****) I$  \\ \hline
$a_f=0.1$  & $0.378054{-}3.08729E{-}11$ I  & $0.394708{-}4.92541E{-}11 I$    & $0.394708{-}5.52827E{-}11 I$  \\ \hline
$a_f=0.15$ & $0.559017{-}7.26757E{-9}$ I &  $0.582304{-}1.60485E{-}8 I$  &  $0.582303{-}2.46023E{-}8I$ \\ \hline
$a_f=0.2$  & $0.730548{-}3.7315E{-}7$ I & $0.758577{-}8.83296E{-}7$ I &  $0.758577{-}9.40976E{-}7 I$  \\ \hline
$a_f=0.25$ & $0.88991{-}7.99721E{-}6$ I & $0.920289{-}0.0000176402$ I  &   $0.920289{-}0.0000181988 I$   \\ \hline
\end{tabular}
\caption{Long-lived modes: $l=2\Rightarrow m_\phi=-1, m_\psi=\pm 1$}
\end{table}

\vspace{1cm}

\begin{table}[h!]
\centering
\begin{tabular}{|c|c|c|}
\hline
$n=0$      & WKB & Seiberg-Witten   \\ \hline
$a_f=0$ (BH) &  1.47902-0.25 I  &   1.50114 - 0.250172 I\\\hline
$a_f=0.01$ & 1.48405 - 0.249139 I   &  1.50378 - 0.247486 I     \\ \hline
$a_f=0.02$ & 1.48914 - 0.248246 I   & 1.50906 - 0.246522 I   \\ \hline
$a_f=0.05$ &1.50476 - 0.245359 I  &  1.52514 - 0.243616 I   \\ \hline
$a_f=0.1$  & 1.53208 - 0.239781 I  &  1.55263 - 0.238765 I   \\ \hline
$a_f=0.15$ & 1.56113 - 0.233085 I & 1.58057 - 0.23374 I     \\ \hline
$a_f=0.2$  & 1.59211 - 0.225054 I  &  1.60861 - 0.227994 I   \\ \hline
$a_f=0.25$ & 1.62519 - 0.215449 I  &  1.63688 - 0.220948 I   \\ \hline
$a_f=0.3$  & 1.66056 - 0.20407 I & 1.66572 - 0.212378 I     \\ \hline
\end{tabular}
\caption{Prompt Ringdown modes: $l=2\Rightarrow m_\phi=2, m_\psi=0$}
\end{table}

\begin{table}
\centering
\begin{tabular}{|c|c|c|}
\hline
$n=0$      & WKB & Seiberg-Witten   \\ \hline
$a_f=0$ (BH) &  1.47902-0.25 I  &   1.50114 - 0.250172 I\\\hline
$a_f=0.01$ & 1.47405 - 0.25083 I    &  1.49339 - 0.249399 I     \\ \hline
$a_f=0.02$ & 1.46913 - 0.251629 I   &  1.48826 - 0.250345 I     \\ \hline
$a_f=0.05$ &1.45471 - 0.253855 I  & 1.4732 - 0.253122 I       \\ \hline
$a_f=0.1$  & 1.43172 - 0.257048 I  &  1.44913 - 0.257473 I      \\ \hline
$a_f=0.15$ & 1.40991 - 0.259679 I  & 1.42631 - 0.261374 I      \\ \hline
$a_f=0.2$  & 1.38921 - 0.26183 I  & 1.40466 - 0.26475 I      \\ \hline
$a_f=0.25$ & 1.36951 - 0.263571 I & 1.3841 - 0.267553 I       \\ \hline
$a_f=0.3$  & 1.35075 - 0.264959 I  & 1.3645 - 0.269752 I     \\ \hline
\end{tabular}
\caption{Prompt Ringdown modes: $l=2\Rightarrow m_\phi=-2, m_\psi=0$}
\end{table}
\FloatBarrier

\end{document}